\documentclass[aps,prl,reprint,superscriptaddress]{revtex4-2}

\usepackage{amsmath}
\usepackage{graphicx} 
\usepackage{gensymb}
\usepackage{siunitx}
\usepackage{soul}
\usepackage{color}

\graphicspath{{figures/}}

\begin{document}


\title{Elasto-capillary network model of inhalation}



\author{Jean-Fran\c cois Louf}
\thanks{These authors contributed equally.}
\affiliation{Department of Chemical and Biological Engineering, Princeton University, Princeton, NJ 08544}
\author{Felix Kratz }
\thanks{These authors contributed equally.}
\affiliation{Department of Chemical and Biological Engineering, Princeton University, Princeton, NJ 08544}
\affiliation{Department of Physics, Technical University of Dortmund, Dortmund, Germany}
\author{Sujit S. Datta}
\email[To whom correspondence should be addressed:\\]{ssdatta@princeton.edu}
\affiliation{Department of Chemical and Biological Engineering, Princeton University, Princeton, NJ 08544}


\date{\today}

\begin{abstract}
The seemingly simple process of inhalation relies on a complex interplay between muscular contraction in the thorax, elasto-capillary interactions in individual lung branches, propagation of air between different connected branches, and overall air flow into the lungs. These processes occur over considerably different length and time scales; consequently, linking them to the biomechanical properties of the lungs, and quantifying how they together control the spatiotemporal features of inhalation, remains a challenge. We address this challenge by developing a computational model of the lungs as a hierarchical, branched network of connected liquid-lined flexible cylinders coupled to a viscoelastic thoracic cavity. Each branch opens at a rate and a pressure that is determined by input biomechanical parameters, enabling us to test the influence of changes in the mechanical properties of lung tissues and secretions on inhalation dynamics. By summing the dynamics of all the branches, we quantify the evolution of overall lung pressure and volume during inhalation, reproducing the shape of measured breathing curves. Using this model, we demonstrate how changes in lung muscle contraction, mucus viscosity and surface tension, and airway wall stiffness---characteristic of many respiratory diseases, including those arising from COVID-19, cystic fibrosis, chronic obstructive pulmonary disease, asthma, and emphysema---drastically alter inhaled lung capacity and breathing duration. Our work therefore helps to identify the key factors that control breathing dynamics, and provides a way to quantify how disease-induced changes in these factors lead to respiratory distress.
\end{abstract}


\maketitle

\section{Introduction}

The ongoing COVID-19 crisis highlights the critical importance of lung biomechanics in our everyday lives: COVID-19 patients frequently develop shortness of breath and often, debilitating and possibly fatal respiratory failure \cite{xu2020pathological,sohrabi2020world,qin2020dysregulation, poyiadji2020covid, chen2020sars}. These complications are thought to arise in part from virus-induced alterations in the biomechanical properties of the lungs---specifically, an increase in the surface tension of the airway mucus lining and a decrease in the strength of the thoracic muscles \cite{xu2020clinical, luo2020clinical}. Such complications also manifest in diverse other disorders arising from cystic fibrosis (CF), chronic obstructive pulmonary disease (COPD), asthma, and emphysema; these are again thought to be linked to changes in airway surface tension or muscular contraction, as well as to other alterations in the mechanics of airway tissues and secretions such as an increase in mucus viscosity and a decrease in airway wall stiffness \cite{polin2014surfactant, mac2016acute, thompson2017acute, oliveira2016entropy}. As a result, treatments frequently rely on mechanical ventilation and exogenous administration of surfactant and/or mucus-thinning agents \cite{poyiadji2020covid, robinson2002mucociliary, newhouse1998intrapulmonary, davis1978assisted, fauroux1999chest, mcelvaney1991aerosol, griese1997nebulization, devendra2002lung, clark2003potential, griese1997pulmonary, homnick1995comparison, holland2003non, hodson1991non}. However, these interventions often proceed by trial-and-error due to a limited understanding of how biomechanical factors impact the overall dynamics of breathing. 

While experiments provide a wealth of information quantifying muscle strength, mucus surface tension and viscosity, and lung airway wall stiffness, directly connecting alterations in these tissue-scale biomechanical factors to organ-scale alterations in breathing is challenging. In particular, measurements of tissue properties can be invasive and often do not provide a way to assess the larger-scale impact of variations in these factors, while measurements of overall breathing dynamics are non-invasive but do not shed light on the underlying biomechanical factors at play. Computational models provide a promising way to overcome these limitations. For example, computational fluid dynamics approaches are capable of resolving air pressure and flow-induced stresses in the lungs with exquisite detail \cite{walters2011efficient, soni2013large, ma2006anatomically, wall2010towards, lewis2005quantification, nowak2003computational, sznitman2007cfd, walters2011computational, walters2010method, kunz2009progress, calay2002numerical, malve2013cfd, wall2008fluid, gemci2008computational}; however, they are computationally intensive and frequently focus on static lung morphologies for simplicity. Conversely, sophisticated pulmonary mechanics models have been developed to elaborate the competition between capillary, viscous, and elastic stresses in determining how individual lung branches deform \cite{hazel2003, heil2008, grotberg2011respiratory, halpern1992fluid, grotberg1994pulmonary, grotberg2004biofluid, heil2011fluid, gaver1996steady, hazel2008influence, juel_heap_2007, heil1999airway, hazel2005surface, heil1997stokes, heil1996large}; however, these models do not incorporate the complex hierarchical structure of the lungs and thus cannot reproduce the full dynamics of breathing. Models that simplify the representation of the different lung branches as an interconnected network provide a promising way to bridge these extremes; however, previous implementations have not treated dynamic changes in lung structure during breathing or have only been used to specifically investigate the influence of structural heterogeneity on breathing \cite{wongviriyawong2010dynamics,politi2010multiscale,donovan2017inter,stewart2015patterns}. Thus, an understanding of how lung biomechanics impacts respiration in general remains elusive.

\begin{figure*}[t]
      \centering
         {\includegraphics[width=\textwidth]{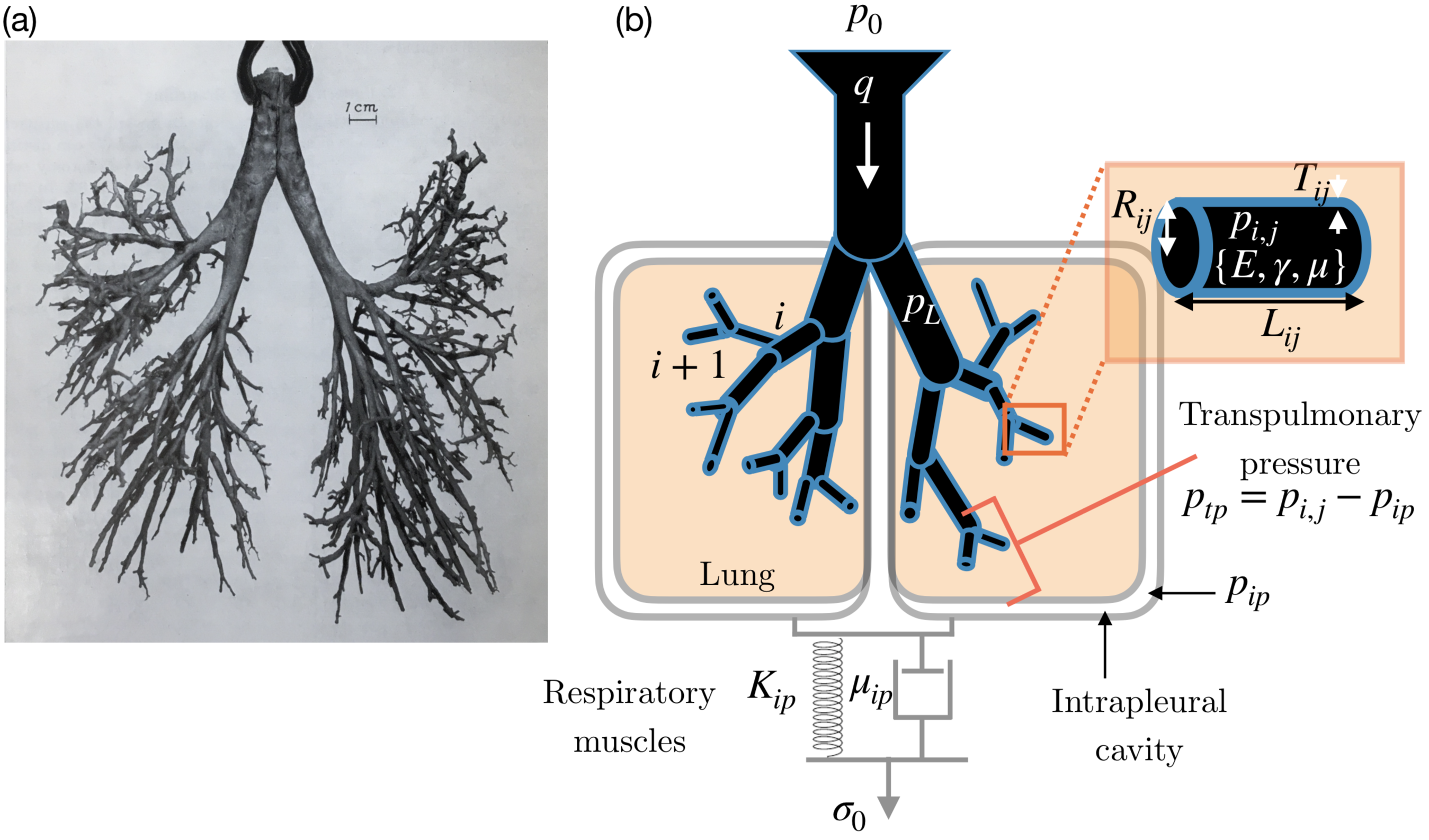}}\\
      \caption{(a) Cast of the bronchial tree of an adult human lung, from \cite{weibel1963}. (b) Schematic of our model of the lungs as a branched network of mucus-lined flexible thin-walled cylinders coupled to a viscoelastic thoracic cavity, represented by the spring and dashpot. Each branch opens at a rate and a pressure that is determined by input biomechanical parameters, enabling us to elucidate how the mechanical properties of lung tissues and secretions impact breathing dynamics.  }\label{Fig1}
 \end{figure*}

Here, we address this problem by developing a dynamic network model of the lungs that connects the multi-scaled processes underlying inhalation: contraction of the thoracic muscles, opening of the individual lung branches, flow of the mucus lining, propagation of air between different connected branches, and overall air flow into the lungs. We hypothesize that the network representation of these processes resolves the relevant length and time scales, while still providing a simplified and computationally tractable representation of the interconnected and hierarchical geometry of the lungs. In support of this hypothesis, we show that our model can describe the evolution of overall lung pressure and volume, as well as the hierarchical and heterogeneous opening of different lung branches, during inhalation starting from a completely closed respiratory zone as a proof of principle. We use the model to demonstrate how disease-induced weakening of the thoracic muscles, increased mucus viscosity and surface tension, and alterations in lung airway wall elasticity impact inhalation. Thus, our results help elucidate how lung biomechanics control breathing dynamics.

\section*{Theory of inhalation dynamics}
\subsection{Lung network representation} Motivated by morphological data, we computationally represent the lungs as a binary branched network of thin-walled, liquid-lined flexible cylinders coupled to a viscoelastic thoracic cavity (Figure \ref{Fig1}). This tree can be classified into two sections leading from the trachea \cite{horsfield1971models, nucci2002morphometric}, indexed by the generation number $i$:  the conducting zone ($0 \leq i \leq 16$), which has a constant open volume and does not contribute to oxygen uptake into the bloodstream, and the respiratory zone ($17 \leq i \leq 23$), which has branches that can collapse and open during respiration, and is the primary site of oxygen uptake. We therefore represent the conducting zone as one static airway branch, and the respiratory zone as a binary tree spanning generations 17 through 23. We index each individual branch by the labels $i$ and $j$, where $17 \leq i \leq 23$ denotes the generation number and $1 \leq j \leq 2^{i}$ corresponds to the index of the branch within a given generation $i$. The branches are all connected; thus, branches deeper in the lungs are only able to open when branches above them have opened. 

Each branch is characterized by an open inner radius $R_{ij}$, length $L_{ij}$, and wall thickness $T_{ij}$, and therefore an open airway volume $V_{ij}=\pi R_{ij}^2L_{ij}$ (Fig. \ref{Fig1}b, right inset). For each generation, the mean values of these morphological parameters $R_{i}\equiv\langle R_{ij} \rangle_j = \sum_{j=1}^{2^{i}} R_{ij}/2^i$, $L_{i}\equiv\langle L_{ij} \rangle_j = \sum_{j=1}^{2^{i}} L_{ij}/2^i$, and $T_{i}\equiv\langle T_{ij} \rangle_j = \sum_{j=1}^{2^{i}} T_{ij}/2^i$ are given by experimental measurements of the mean branch radius, length, and thickness, respectively (Table \ref{Tab:geometry}). To incorporate heterogeneity, a natural feature of the lungs, we then randomly select the individual $R_{ij}$, $L_{ij}$, and $T_{ij}$ from a uniform distribution bounded by $\pm$ 25 $\%$ of $R_i$, $L_i$, and $T_i$, respectively. The results shown in Figs. \ref{Fig2}--\ref{Fig6} all utilize the same lung architecture parameterized by the same values of $\{R_{ij},L_{ij},T_{ij}\}$, to isolate the influence of biomechanical factors on breathing. However, an advantage of our network representation is that it is generalizable: specific values of the morphological parameters can be incorporated in future extensions of this work. For example, our model could be used to assess the distributions of outcomes across different airways with different $\{R_{i},L_{i},T_{i}\}$, or between different realizations of the same airways having the same $\{R_{i},L_{i},T_{i}\}$, given the importance of structural heterogeneity on breathing \cite{stewart2015patterns}.

\begin{table*}
  \centering
  \begin{tabular}{|c|c|c|c|}
    \hline
  Generation $i$ & Mean radius $R_i$ (mm) & Mean length $L_i$ (mm) & Mean wall thickness $T_i$ (mm)  \\
    \hline
    17 & 0.270 & 1.41 & 0.0236 \\
	18 & 0.250 & 1.17 & 0.0229 \\
	19 & 0.235 & 0.99 & 0.0226 \\
	20 & 0.225 & 0.83 & 0.0227 \\
	21 & 0.215 & 0.70 & 0.0228 \\
	22 & 0.205 & 0.59 & 0.0231 \\
	23 & 0.204 & 0.50 & 0.0250 \\  \hline
\end{tabular}
\caption{Morphological parameters used in our simulations, obtained from experimental measurements \cite{weibel1963, habib1994}.}
  \label{Tab:geometry}
\end{table*}

\begin{table*}
  \centering
  \begin{tabular}{|l|c|c|c|}
    \hline
  Biomechanical parameters & Value & Reference  \\
    \hline
    Young's modulus of the lung airway wall $E$ & 5 kPa & \cite{wang2011, lewis2001mechanics} \\
    Poisson's ratio of the airway wall $\nu$ & 0.5 & \cite{laifook1978} \\
	Mucus dynamic shear viscosity $\mu$ & 100 mPa-s & \cite{baconnais1999, puchelle1983, matsui2006} \\
	Applied muscular stress $\sigma_0$ & 500 Pa & \cite{hughes1999, bland1986, ker1998, polkey1996} \\
	Initial volume of the intrapleural cavity V$_{ip,0}$ & 20 mL  & \cite{d2019physiology} \\
	Maximal open airway volume V$_0$ & 1.675 L & \cite{weibel1963} \\
	Initial pressure of the intrapleural cavity $p_{ip,0}$ & $p_0$ -- 400 Pa & \cite{christie1934} \\
	Effective bulk modulus of the intrapleural cavity $K_{ip}$ & 100 kPa & \cite{wang2011, lewis2001mechanics} \\
	Mucus surface tension  $\gamma$ & 15 mN/m & \cite{geiser2003, alonso2005} \\ \hline
\end{tabular}
\caption{Biomechanical parameters used in our simulations, obtained from experimental measurements.}
  \label{Tab:physilogy}
\end{table*}

Further, to incorporate the biomechanical properties of lung tissues, we make the simplifying assumption that the inner walls of the branches are uniformly coated by a Newtonian fluid of negligible thickness with dynamic shear viscosity $\mu$ and surface tension $\gamma$, and the lung airway wall is a linear elastic solid with Young's modulus $E$; we use values of $\mu$, $\gamma$, and $E$ obtained from experimental measurements, as listed in Table \ref{Tab:physilogy}, and take them to be constant throughout the lungs. This model therefore represents a key first step toward computationally describing the lungs, which in reality have non-Newtonian mucus with a non-negligible, generation-dependent thickness, as well as generation-dependent values of the parameters $\mu$, $\gamma$, and $E$. However, our network representation enables specific branch-dependent values of these biomechanical parameters to be incorporated, which would be another useful direction for future work.

\subsection{Stress exerted by thoracic muscles} As a first step toward modeling the full dynamics of respiration, here we consider the process of inhalation starting from a completely closed respiratory zone---characteristic of newborn infants or patients with severe respiratory distress \cite{ellwein2018theoretical}. We therefore initialize the model with all branches with $17\leq i\leq23$ closed. Building on this model to explore the additional dynamics of exhalation as well as multiple breathing cycles will be an important direction for future research.

Inhalation begins with the contraction of the thoracic muscles that, as a first approximation, we assume pull with a constant stress $\sigma_0$ on the intrapleural cavity. Motivated by previous work \cite{zhang2013dynamic, lewis2001mechanics}, we model the viscoelastic behavior of the chest using a Kelvin-Voigt model, which treats the chest as a combination of an elastic spring and a viscous dashpot connected in parallel (Fig. \ref{Fig1}b, bottom): $ \sigma(t) \equiv \sigma_0 = K_{ip} \varepsilon(t) + \mu_{ip} \dot{\varepsilon}(t)$, where $K_{ip}$ and $\mu_{ip}$ are the effective elastic and viscous constants characterizing the intrapleural cavity and $\varepsilon\equiv\Delta V_{ip}(t)/V_{ip,0}$ represents the volumetric strain in the intrapleural cavity, where $\Delta V_{ip}$ represents the difference in the cavity volume compared to its initial value $V_{ip,0}$. Thus, the intrapleural space expands over time in a stress-dependent manner: 
\begin{equation}
  \varepsilon(t) =\varepsilon_{max}\left(1-e^{-t/\tau_{B}}\right)
  \label{eqn:epsilon}
\end{equation}
\noindent where $\varepsilon_{max}\equiv\sigma_{0}/K_{ip}$ is the maximal strain and $\tau_{B}\equiv\mu_{ip}/K_{ip}$ is the characteristic time scale of inhalation. Based on previous measurements \cite{wang2011, lewis2001mechanics}, we take $K_{ip}=100$ kPa and $\tau_{B}=1$ s.

\subsection{Elasto-capillary interactions in individual branches} 	The expansion of the thoracic cavity reduces the intrapleural pressure $p_{ip}$ (Fig. \ref{Fig1}, right): given a fixed amount of air within the intrapleural space, $p_{ip}(t)=p_{ip,0}/\left[1+\varepsilon(t)\right]$, where $p_{ip,0}\approx99.6$ kPa as determined experimentally \cite{blom2003} and $\varepsilon(t)$ is given by Eq. \ref{eqn:epsilon}. Thus, the transpulmonary pressure difference across the lung airway wall $p_{tp}(t)\equiv p_{L}(t)-p_{ip}(t)$ transiently increases; here $p_{L}(t)$ is the air pressure in the respiratory zone, which we take to be constant throughout due to the low air flow resistance of the respiratory zone, as justified in the \textit{Methods}. 

As established in many previous studies \cite{suki1994avalanches, suki2000size, bates2002time, stewart2015patterns}, as $p_{tp}$ increases, it exceeds the threshold pressure $p_{ij}^{th}=8\gamma/R_{ij}$ required to open a collapsed branch $ij$ that is in contact with the open region of the lungs \cite{hazel2003}. In this case, an air finger propagates into the branch at a speed $U_{ij}$ that is determined by a complex interplay between viscous forces in the airway mucus lining as the branch is pulled apart, capillary forces holding the walls of the branch together, and elastic forces resisting bending of the branch walls. Motivated by the results of previous three-dimensional numerical solutions \cite{hazel2003}, we estimate this speed using the relation $\mathrm{Ca}_{ij} = \frac{1}{\Gamma_{ij}} \frac{p_{tp}-p_{ij}^{th}}{\gamma / R_{ij}}$ where $\mathrm{Ca}_{ij}\equiv \mu U_{ij}/\gamma$ is the capillary number and $\Gamma_{ij}\equiv\left(\gamma/R_{ij}\right)/B_{ij}$ is a dimensionless parameter that quantifies the competition between capillary and elastic stresses in the branch, with bending stiffness $B_{ij}\equiv E\left(T_{ij}/R_{ij}\right)^{3}\bigg/12\left(1-\nu^2\right)$ and $\nu\approx0.5$ is the Poisson's ratio of the airway wall. This equation highlights three key features of branch opening. First, capillary forces hold the walls of a branch together, and thus, the transpulmonary pressure $p_{tp}$ must overcome the capillary pressure threshold $p_{ij}^{th}$ to force a branch to open. Second, the elastic energy penalty associated with deforming a branch also promotes opening, as quantified by $\Gamma_{ij}$. Third, branch opening is not instantaneous, but is limited by viscous dissipation as the mucus lining is pushed apart, as quantified by $\mathrm{Ca}_{ij}$.


We directly implement this relation into our model, with the condition that a given branch can only open if its proximal end is in contact with the open region of the lungs. For ease of computation we treat the branch as being split into a fully open fraction with time-dependent volume $V_{ij}(t)$ and a remaining fully closed fraction. In non-dimensional form, this relation can then be expressed as
\begin{equation}
\frac{\partial \hat{V}_{ij}}{\partial \hat{t}} = \frac{\zeta \hat{R}_{ij}^3}{\hat{\Gamma}_{ij}}  \left[\hat{p}_{tp}(\hat{t})-\hat{p}_{ij}^{th}\right]
\label{eq2.8}
\end{equation}
where the hat notation $(\hat{})$ indicates that the variables $V_{ij}$, $t$, $R_{ij}$, $\Gamma_{ij}$, and $p_{tp}-p_{ij}^{th}$ have been normalized by the characteristic branch-scale quantities $\pi R_{17}^{2}L_{17}$, $\mu/p_{17}^{th}$, $R_{17}$, $\left(\gamma/R_{17}\right)/B_{17}$, and $p_{17}^{th}$, respectively, where the subscript 17 refers to the mean value at the first generation of the respiratory zone. This non-dimensional form reveals that the biomechanical parameter $\zeta\equiv(R_{17}/L_{17})/\Gamma_{17}$ is a key factor that governs the amount of lung opening during inhalation: when $\zeta$ is large, elasticity tends to peel the lung branches open, while when $\zeta$ is small, capillarity tends to hold lung branches shut. 

\begin{figure*}[t]
     \centering
         {\includegraphics[width=\textwidth]{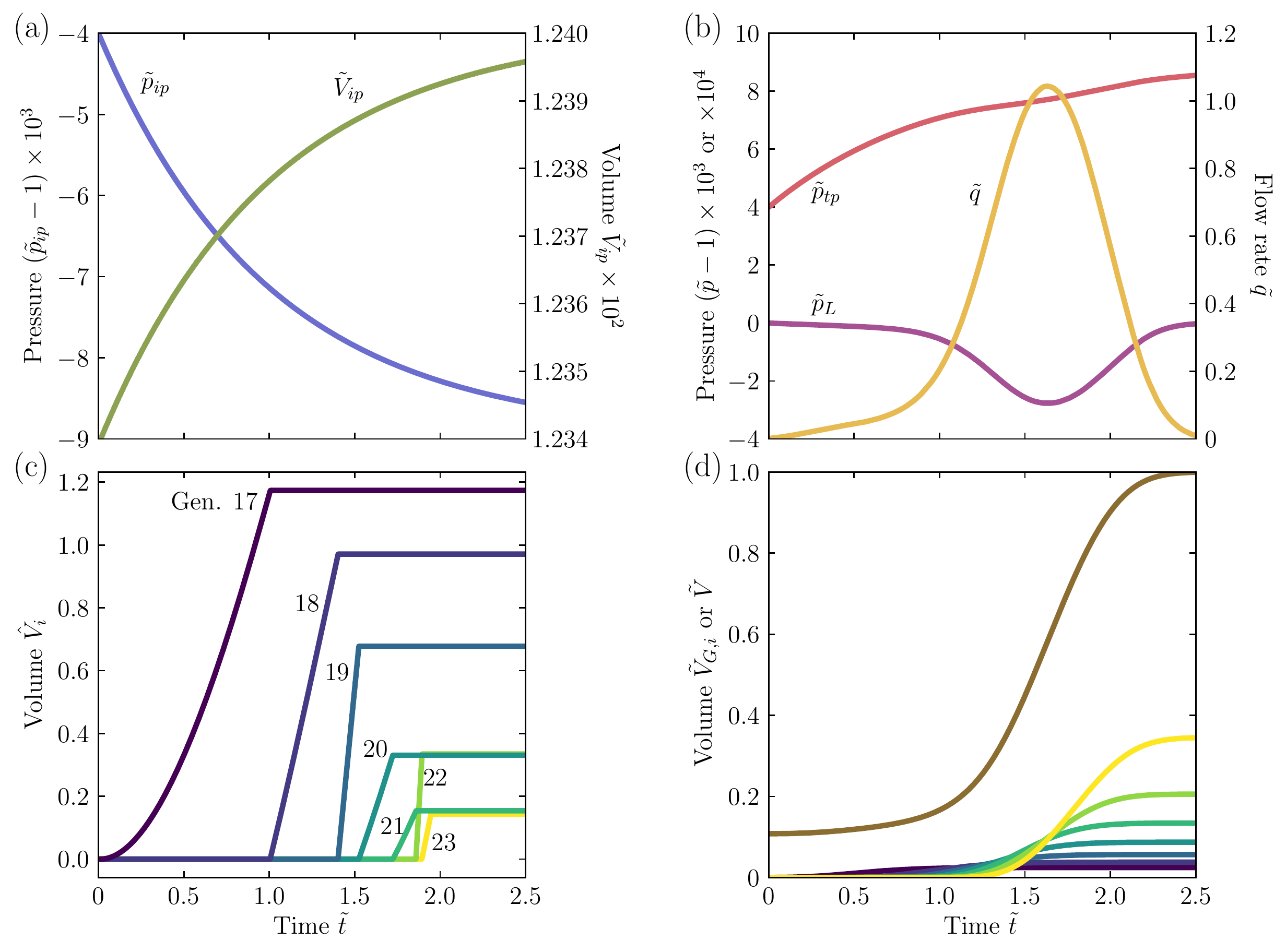}}\\
      \caption{Typical inhalation dynamics. Plots show the simulated evolution of lung pressures and volumes as a function of the dimensionless inhalation duration $\tilde{\tau}\equiv t/\tau_B$, where $\tau_B=1$ s is the characteristic inhalation time. (a) Pressure $p_{ip}$ (blue) and volume $V_{ip}$ (green) of the intrapleural cavity, normalized by atmospheric pressure $p_0$ and the maximal lung volume $V_0$, respectively. To show the small variation of $\tilde{p}_{ip}$ and $\tilde{V}_{ip}$ more clearly, we plot $(\tilde{p}_{ip}-1)\times10^{3}$ (left) and $\tilde{V}_{ip}\times10^{2}$ (right). (b) Transpulmonary pressure $p_{tp}$ (red), respiratory zone pressure $p_{L}$ (purple), and air inflow rate $q$ (yellow). Pressures are normalized by atmospheric pressure while flow rate is normalized by the characteristic flow rate $V_0/\tau_B$. To show the small variation of the pressures more clearly, we plot $(\tilde{p}-1)\times10^{3}$ in the case of $p=p_{tp}$ and $(\tilde{p}_{L}-1)\times10^{4}$ in the case of $p=p_{L}$. (c) Open volumes of individual connected branches in each generation of the respiratory zone, normalized by the characteristic branch-scale volume $\pi R_{17}^{2}L_{17}$, where $R_{17}$ and $L_{17}$ are the mean values of the branch radius and length at the first generation of the respiratory zone, respectively; different numbers by the curves indicate the generation number $i$. (d) Dark blue to yellow curves show the total open volume summed over each generation $i$, $V_{G,i}$, with the colors corresponding to the same generations as in panel (c). The brown curve shows the total open volume of airways of the lung, $V$. Both $V_{G,i}$ and $V$ are normalized by the maximal lung airway volume. In all these simulations, applied muscular stress $\sigma_0=500$ Pa, mucus viscosity $\mu=100$ mPa-s, mucus surface tension $\gamma=15$ mN/m, and biomechanical parameter $\zeta=1.3\times10^{-3}$.
}\label{Fig2}
 \end{figure*}

\subsection{Overall opening of the lungs} The physics described in the previous subsection governs the opening of individual branches; summing over all open regions of the airways then yields the total opened lung volume $V(t)=V_{C}+V_{R}(t)=V_{C}+\sum_{i=17}^{23}\sum_{j=1}^{2^{i}}V_{ij}(t)$, where $V_{C}$ is the constant open volume of the conducting zone and $V_{R}$ is the time-dependent open volume of the respiratory zone, with $V_{ij}$ given by Eq. \ref{eq2.8}. Thus, as $p_{tp}$ increases, the open volume of the respiratory zone $V_{R}$ increases, causing the lung pressure $p_{L}$ to transiently decrease. This decrease in pressure draws air into the lungs from the atmosphere with a volumetric flow rate of magnitude $q(t)\sim \left[p_0 - p_L(t)\right] \bigg/\sum_{i=0}^{16} \left( \sum_{j=1}^{2^{i}} \Omega_{ij}^{-1} \right)^{-1} $ through the conducting zone (\textit{Methods}), where $p_{0}\approx 101$ kPa is atmospheric pressure.

\subsection{Computational implementation of the model}

Analysis of these coupled processes enables us to quantitatively model the full dynamics of inhalation. Specifically, as detailed in the \textit{Methods}, we input values of the morphological parameters $\{ R_{ij},L_{ij},T_{ij} \}$ and the biomechanical parameters $\{ \mu,\gamma,E,\sigma_{0} \}$ and iteratively solve the equations described in the \textit{Theory} section at uniform discrete time steps $\Delta t$. This scheme thus enables us to determine the full evolution of the pressures $\{\tilde{p}_{ip},\tilde{p}_{tp},\tilde{p}_{L}\}$, the volumes $\{\tilde{V}_{ip},\hat{V}_{ij},\tilde{V}\}$ and the flow rate $\tilde{q}$ over time $\tilde{t}$, where the tilde notation $(\tilde{})$ indicates lung-scale variables that have been normalized by the atmospheric pressure $p_{0}$, maximal open lung volume $V_{0}$, characteristic flow rate $V_{0}/\tau_{B}$, and breathing time $\tau_{B}$ respectively. Importantly, this network representation reduces computational cost: the complete dynamics of inhalation can be obtained within a matter of minutes on a conventional personal computer. For example, the entirety of the results in Fig. \ref{Fig2} were obtained using a laptop with an 8th Gen Intel Core i7-8750H 6 core processor with 2.2 GHz and 16GB RAM in 20 minutes. Thus, our network model provides a computationally tractable way to characterize the dynamics of respiration that can be implemented by specialists and general users alike.



\section*{Results}
\subsection*{Typical inhalation dynamics} We begin by describing the full inhalation dynamics of a representative healthy lung, using published measurements for the input parameters \cite{weibel1963,geiser2003,alonso2005,wang2011,lewis2001mechanics,baconnais1999,puchelle1983,matsui2006,hughes1999,bland1986,ker1998,polkey1996,d2019physiology,christie1934}.  The simulation captures the expected dynamics of inhalation. First, the contraction of the respiratory muscles generates a stress on the intrapleural cavity, expanding it (Fig. \ref{Fig2}a, green) and reducing its internal pressure (Fig. \ref{Fig2}a, blue). The transpulmonary pressure difference between the lung interior and the intrapleural space subsequently builds up (Fig. \ref{Fig2}b, red), causing respiratory branches to successively open (Fig. \ref{Fig2}c-d, dark blue to green to yellow), reducing the pressure in the respiratory zone (Fig. \ref{Fig2}b, purple) and driving air flow into the lungs (Fig. \ref{Fig2}b, yellow). This process continues as the intrapleural cavity expands over time. Eventually, however, the applied stress is able to expand the chest by less and less (Fig. \ref{Fig2}a, plateau in green curve), and branches open at a slower rate; the flow rate of air into the lungs eventually reaches zero at a time $\tilde{t}_{max} \approx 2.5$ (Fig. \ref{Fig2}b, yellow), and inhalation ceases. 

Each generation is made of progressively smaller branches having progressively larger threshold pressures for opening. Thus, we expect lung opening to be hierarchical: the larger proximal branches open first, and then smaller distal branches open later, after air is able to propagate to them and exceed the capillary pressure threshold. Our computational model captures this hierarchy of branch opening, as shown in Figs. \ref{Fig2}c-d, complementing previous investigations of collective branch opening \cite{suki1994avalanches, suki2000size, bates2002time, stewart2015patterns}. Fig. \ref{Fig2}c shows an example of individual, connected branches in different generations as they open. The branch in the first generation of the respiratory zone opens first ($i = 17$, dark blue); once it has fully opened, air can propagate into the branch in the next generation ($i = 18$, lighter blue), causing successive opening of branches through the different generations and eventually reaching the terminal alveoli ($i = 23$, yellow). Notably, however, even though these terminal generations must open later, and though their individual branches are smaller, they collectively contribute the largest volume to the open lung, as shown in Fig. \ref{Fig2}d; the dark blue to yellow curves show the volume of all the open branches in a given generation $i$, $V_{G,i}(t)=\sum_{j=1}^{2^{i}}V_{ij}(t)$, while the brown curve shows the total open volume of the lung $V(t)$. Though the first generation of the respiratory zone (dark blue) is the first to open during inhalation, it contributes only $\sim 3\%$ volume to the open lung; by contrast, the terminal generation (yellow) is the last to open, but contributes $\sim 35\%$ of the overall lung volume. Thus, our model quantifies the expectation that opening later generations is key for healthy lung performance.

\subsection*{Influence of changes in biomechanical parameters on inhalation}

Having characterized the typical dynamics of inhalation, we next investigate how these dynamics are controlled by key biomechanical factors. Specifically, motivated by their relevance to respiratory distress stemming from the ongoing COVID-19 crisis, as well as from prevalent conditions such as CF, COPD, asthma, and emphysema, we focus on the role of four key factors: (i) Muscle-induced stress $\sigma_0$, (ii) Mucus viscosity $\mu$, (iii) Mucus surface tension $\gamma$, and (iv) Airway wall stiffness $E$. 

\begin{figure*}
     \centering
         {\includegraphics[width=\textwidth]{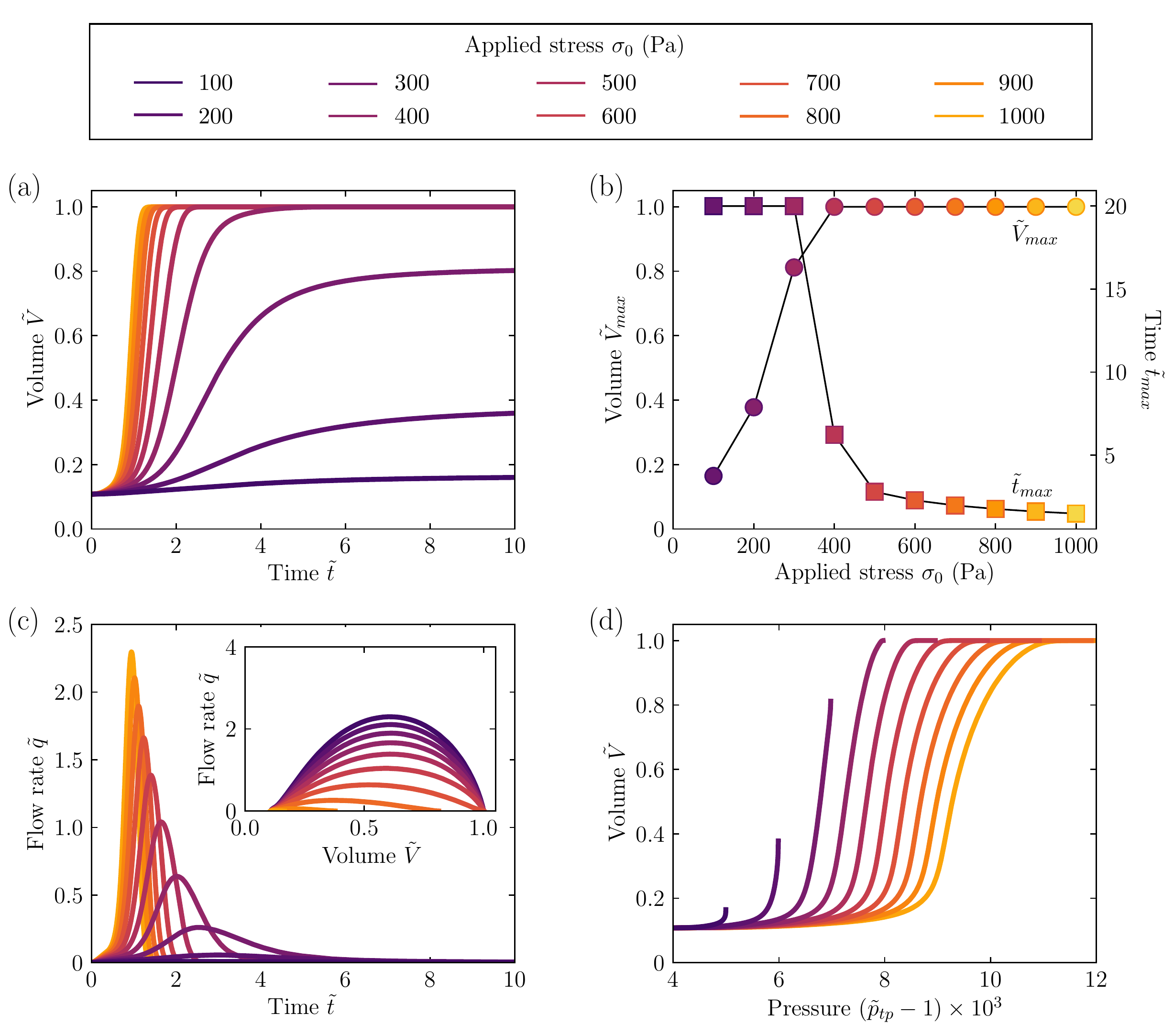}}\\
      \caption{Stress applied by thoracic muscles strongly regulates dynamics and full extent of inhalation. In all panels, different colors show different applied stresses $\sigma_{0}$. (a) Simulated evolution of the total open volume of the lung airways $V$, normalized by the maximal lung volume $V_0$, as a function of the dimensionless inhalation duration $\tilde{\tau}\equiv t/\tau_B$, where $\tau_B=1$ s is the characteristic inhalation time. (b) Maximal opened volume of the lung airways $V_{max}$ (circles) decreases, and the time needed to reach this volume $t_{max}$ (squares) increases, with decreasing applied stress. Volume and time are normalized by the maximal lung volume $V_0$ and the characteristic inhalation time $\tau_{B}$, respectively. Simulations are performed for times up to $\tilde{t}=20$; therefore, values of $\tilde{t}_{max}$ and $\tilde{V}_{max}$ for $\sigma_{0}\leq300$ Pa are truncated, and could be even larger for simulations run over longer durations. (c) Air inflow rate $q$, normalized by the characteristic flow rate $V_0/\tau_B$, over time. Inset shows the variation of the flow rate with total opened lung volume, as is often measured experimentally via spirometry. (d) Opened lung volume $V$-transpulmonary pressure $p_{tp}$ curves, as is often measured experimentally. Pressure is normalized by atmospheric pressure $p_{0}$; to show the small variation of $\tilde{p}_{tp}$ more clearly, we plot $(\tilde{p}_{tp}-1)\times10^{3}$ on the horizontal axis. In all these simulations, mucus viscosity $\mu=100$ mPa-s, mucus surface tension $\gamma=15$ mN/m, and biomechanical parameter $\zeta=1.3\times10^{-3}$.}\label{Fig_sigma}
 \end{figure*}

\subsubsection*{Role of muscle-induced stress} Patients with COVID-19, CF, or COPD frequently exhibit fatigue and muscle weakness \cite{wang2020updated, kim2013chronic}; similar symptoms also manifest in patients who have undergone mechanical ventilation as a treatment for prolonged periods of time \cite{tobin2010narrative}. The analysis presented in the \textit{Theory} section suggests that this decrease in $\sigma_0$ reduces the expansion of the intrapleural cavity during inhalation, limiting the amount of air that can be taken into the lungs and giving rise to respiratory distress.

Our simulations with varying $\sigma_{0}$ confirm this expectation. In particular, we find that the dynamics of lung opening strongly depend on the applied stress (Fig. \ref{Fig_sigma}a), indicating that it is a key regulator of breathing; reducing the stress exerted by the thoracic muscles decreases the rate at which air is drawn in and prolongs the overall duration of inhalation (Fig. \ref{Fig_sigma}c). Indeed, when $\sigma_0$ is reduced to just half of its typical healthy value $\approx500$ Pa \cite{hughes1999, bland1986, ker1998, polkey1996}, the full duration of lung opening takes nearly ten times longer, and only reaches half the fully opened volume, as shown by the squares and circles in Fig. \ref{Fig_sigma}b, respectively. The corresponding stress-dependent pressure-volume (Fig. \ref{Fig_sigma}d) and flow rate-volume (Fig. \ref{Fig_sigma}c, inset) curves obtained in our simulations are strikingly similar to those observed in experimental measurements \cite{kozlowska2005spirometry}. Thus, our computational approach provides a way to quantify the impact of specific changes in muscle-induced stress on inhalation, shedding light on its relative influence in causing respiratory distress.

\begin{figure*}
     \centering
         {\includegraphics[width=\textwidth]{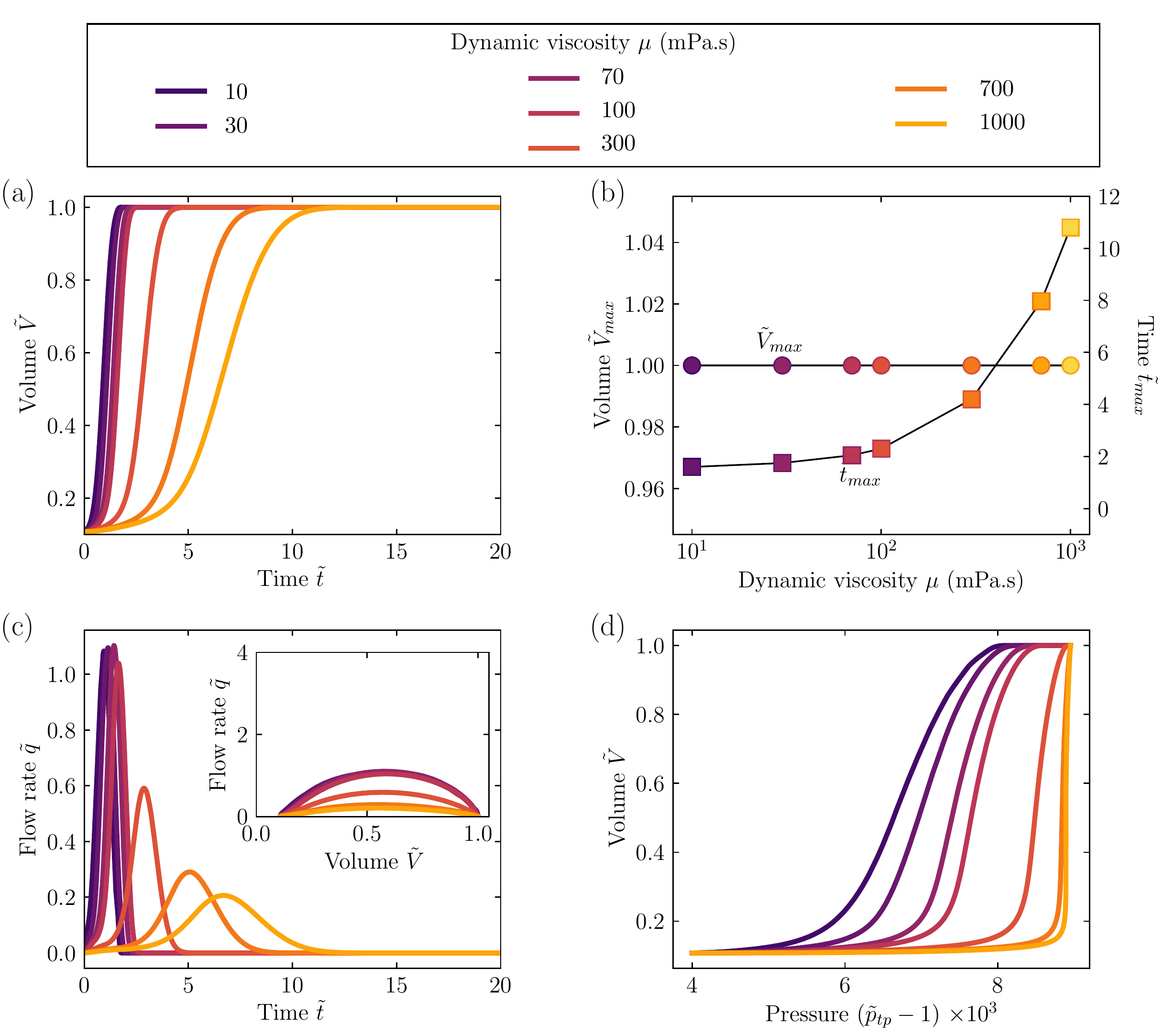}}\\
      \caption{Mucus viscosity strongly regulates inhalation dynamics. In all panels, different colors show different surface tensions $\mu$. (a) Simulated evolution of the total open volume of the lung airways $V$, normalized by the maximal lung volume $V_0$, as a function of the dimensionless inhalation duration $\tilde{\tau}\equiv t/\tau_B$, where $\tau_B=1$ s is the characteristic inhalation time. (b) Maximal opened volume of the lung airways $V_{max}$ (circles) is constant, but the time needed to reach this volume $t_{max}$ (squares) increases with increasing viscosity: mucus viscosity acts as a time scaling parameter. Volume and time are normalized by the maximal lung volume $V_0$ and the characteristic inhalation time $\tau_{B}$, respectively. (c) Air inflow rate $q$, normalized by the characteristic flow rate $V_0/\tau_B$, over time. Inset shows the variation of the flow rate with total opened lung volume, as is often measured experimentally via spirometry. (d) Opened lung volume $V$-transpulmonary pressure $p_{tp}$ curves, as is often measured experimentally. Pressure is normalized by atmospheric pressure $p_{0}$; to show the small variation of $\tilde{p}_{tp}$ more clearly, we plot $(\tilde{p}_{tp}-1)\times10^{3}$ on the horizontal axis. In all these simulations, applied muscular stress $\sigma_{0}=500$ Pa, mucus surface tension $\gamma=15$ mN/m, and biomechanical parameter $\zeta=1.3\times10^{-3}$.
      }\label{Fig_mu}
 \end{figure*}

 \begin{figure*}
     \centering
         {\includegraphics[width=\textwidth]{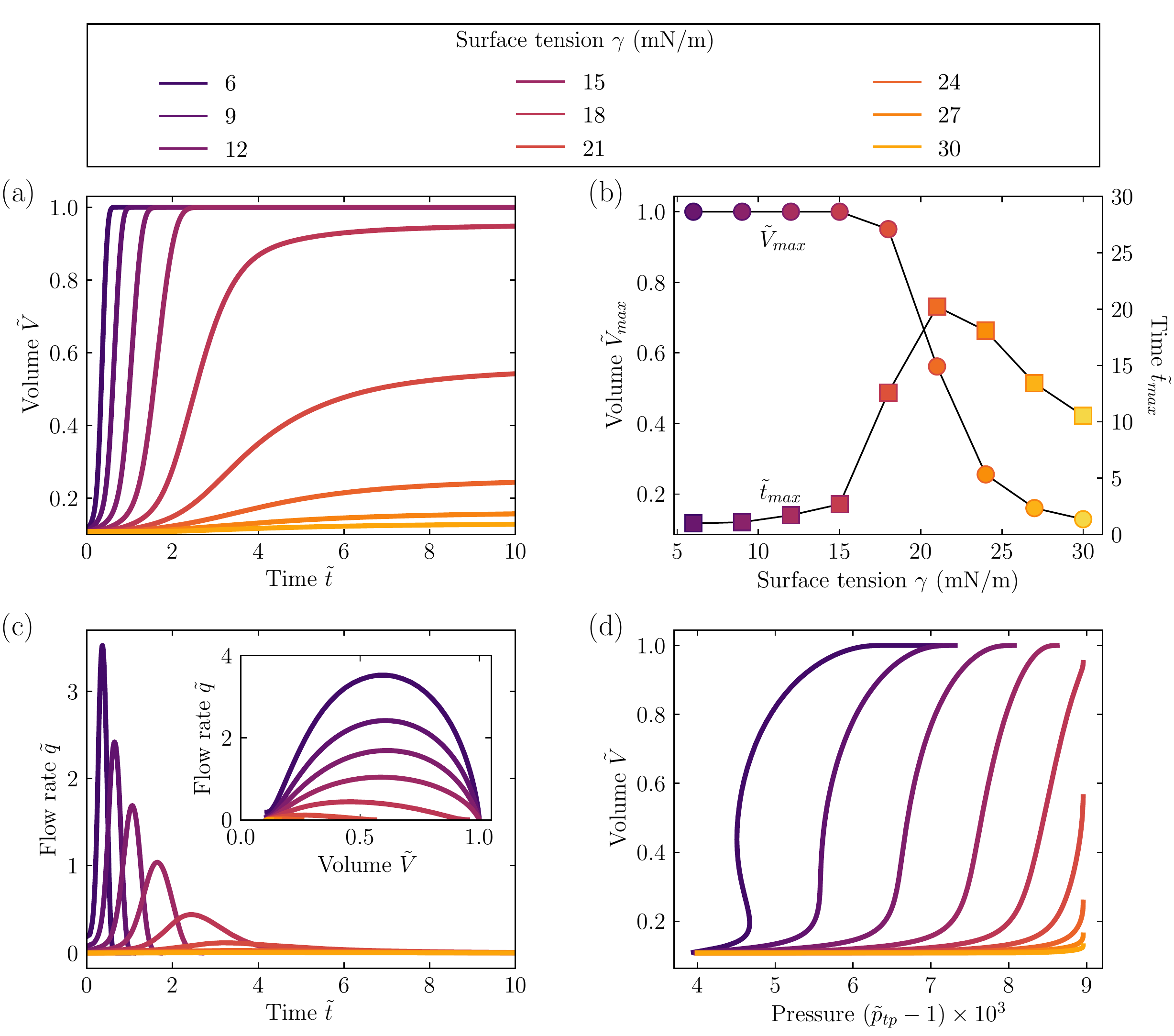}}\\
           \caption{Mucus surface tension strongly regulates full extent and dynamics of inhalation. In all panels, different colors show different surface tensions $\gamma$. (a) Simulated evolution of the total open volume of the lung airways $V$, normalized by the maximal lung volume $V_0$, as a function of the dimensionless inhalation duration $\tilde{\tau}\equiv t/\tau_B$, where $\tau_B=1$ s is the characteristic inhalation time. (b) Maximal opened volume of the lung airways $V_{max}$ (circles) decreases, and the time needed to reach this volume $t_{max}$ (squares) increases and then decreases, with increasing surface tension. Volume and time are normalized by the maximal lung volume $V_0$ and the characteristic inhalation time $\tau_{B}$, respectively. (c) Air inflow rate $q$, normalized by the characteristic flow rate $V_0/\tau_B$, over time. Inset shows the variation of the flow rate with total opened lung volume, as is often measured experimentally via spirometry. (d) Opened lung volume $V$-transpulmonary pressure $p_{tp}$ curves, as is often measured experimentally. Pressure is normalized by atmospheric pressure $p_{0}$; to show the small variation of $\tilde{p}_{tp}$ more clearly, we plot $(\tilde{p}_{tp}-1)\times10^{3}$ on the horizontal axis. In all these simulations, applied muscular stress $\sigma_{0}=500$ Pa, mucus viscosity $\mu=100$ mPa-s, and biomechanical parameter $\zeta=1.3\times10^{-3}$.}\label{Fig_gamma}
 \end{figure*} 
 
   \begin{figure*}
     \centering
         {\includegraphics[width=\textwidth]{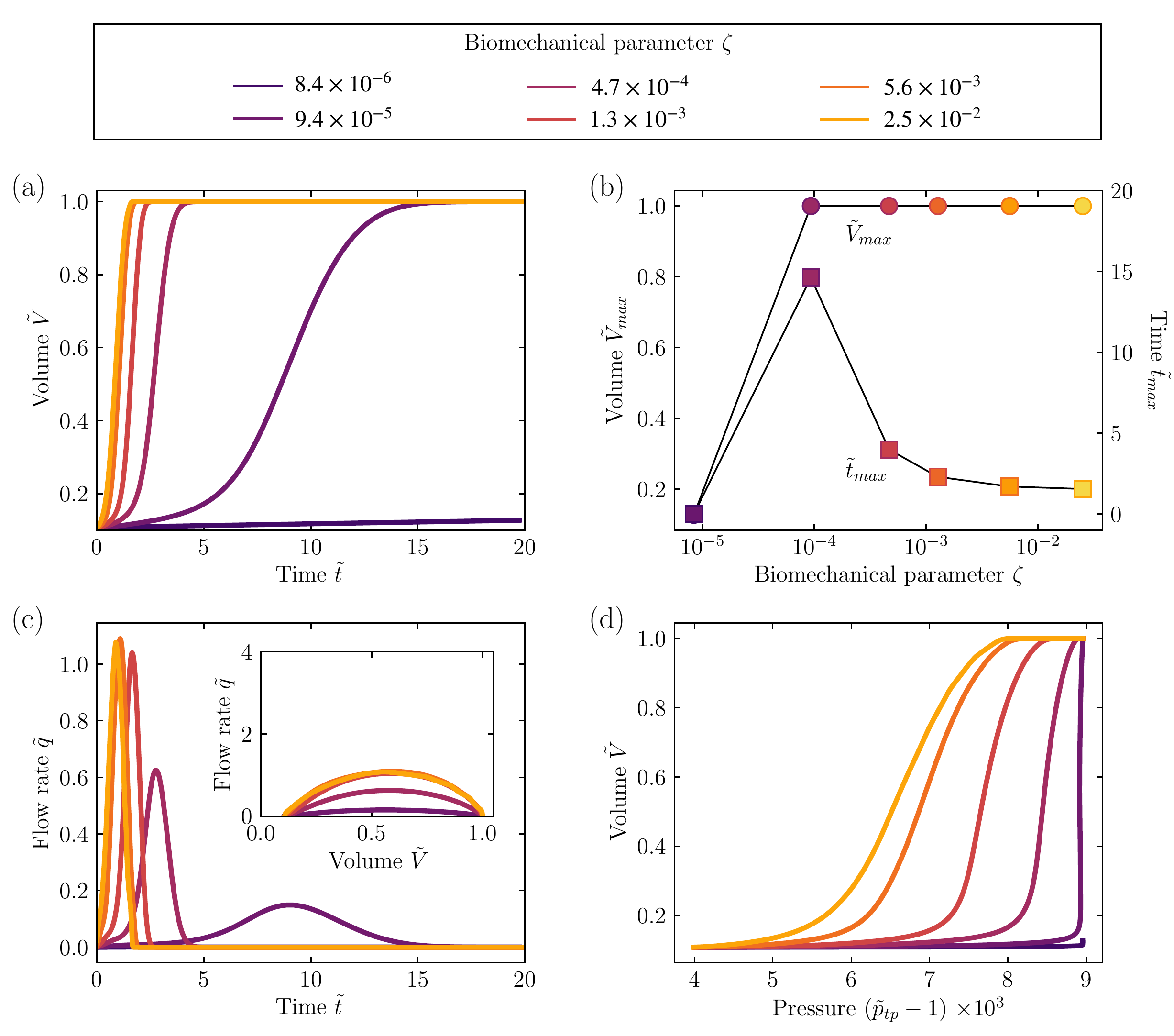}}\\
         \caption{Competition between elasticity and capillarity, quantified by the biomechanical parameter $\zeta$, strongly regulates full extent and dynamics of inhalation. In all panels, different colors show different $\zeta$. (a) Simulated evolution of the total open volume of the lung airways $V$, normalized by the maximal lung volume $V_0$, as a function of the dimensionless inhalation duration $\tilde{\tau}\equiv t/\tau_B$, where $\tau_B=1$ s is the characteristic inhalation time. (b) Maximal opened volume of the lung airways $V_{max}$ (circles) decreases, and the time needed to reach this volume $t_{max}$ (squares) increases and then decreases, with increasing $\zeta$. Volume and time are normalized by the maximal lung volume $V_0$ and the characteristic inhalation time $\tau_{B}$, respectively. (c) Air inflow rate $q$, normalized by the characteristic flow rate $V_0/\tau_B$, over time. Inset shows the variation of the flow rate with total opened lung volume, as is often measured experimentally via spirometry. (d) Opened lung volume $V$-transpulmonary pressure $p_{tp}$ curves, as is often measured experimentally. Pressure is normalized by atmospheric pressure $p_{0}$; to show the small variation of $\tilde{p}_{tp}$ more clearly, we plot $(\tilde{p}_{tp}-1)\times10^{3}$ on the horizontal axis. In all these simulations, applied muscular stress $\sigma_{0}=500$ Pa, mucus viscosity $\mu=100$ mPa-s, and mucus surface tension $\gamma=15$ mN/m.}\label{Fig_zeta}
 \end{figure*}

\subsubsection*{Role of mucus viscosity} 

A common symptom of bronchitis, CF, COPD, interstitial lung disease, and possibly COVID-19 is a large increase in the viscosity of lung mucus \cite{puchelle1983, lai2009micro, rubin2007mucus, sturgess1970viscosity, ye2020chest}. The analysis presented in the \textit{Theory} section suggests that this increase in $\mu$ increases the time scale over which individual lung branches open, possibly slowing the opening dynamics during inhalation and giving rise to respiratory distress.

Our simulations with varying $\mu$ confirm this expectation. In particular, we find that the dynamics of lung opening strongly depend on the mucus viscosity (Fig. \ref{Fig_mu}a), indicating that it is another key regulator of breathing; increasing the mucus viscosity increases the time needed to reach the capillary pressure threshold $p^{th}$ and open airway branches, decreasing the rate at which air is drawn in and prolonging the overall duration of inhalation (Fig. \ref{Fig_mu}c). Indeed, when $\mu$ is increased by a factor of $\sim10$ from its typical healthy value $\approx100$ mPa-s, as can be the case in many lung diseases \cite{puchelle1983, lai2009micro, rubin2007mucus, sturgess1970viscosity, ye2020chest}, the fully opened lung volume is unchanged, but the full duration of lung opening takes approximately five times longer, as shown by the circles and squares in Fig. \ref{Fig_mu}b, respectively. Thus, alterations in mucus viscosity alter the dynamics, but not full extent, of lung opening during inhalation. The corresponding viscosity-dependent pressure-volume (Fig. \ref{Fig_mu}d) and flow rate-volume (Fig. \ref{Fig_mu}c, inset) curves obtained in our simulations are again strikingly similar to those observed in experimental measurements \cite{kozlowska2005spirometry}. Thus, our computational approach provides a way to quantify the impact of specific changes in mucus viscosity on inhalation, shedding light on its relative influence in causing respiratory distress.
 
 \subsubsection*{Role of mucus surface tension} One of the most prominent pathological features of COVID-19 is hindered production of lung surfactant due to viral infection, resulting in a large increase in the surface tension of airway mucus \cite{bracco2020covid, gunther2001surfactant, gralinski2015molecular, yi2020covid}. Similar complications arise in COPD and possibly in asthma and emphysema \cite{kim2013chronic, hohlfeld2001role, ingenito2005role}. The analysis presented in the \textit{Theory} section suggests that this increase in $\gamma$ has two key effects, both of which could contribute to respiratory distress. First, it increases the threshold pressure $p_{ij}^{th}=8\gamma/R_{ij}$ required to open a collapsed branch $ij$. Second, it decreases the biomechanical parameter $\zeta\propto 1/\gamma$, which quantifies the competition between elastic and capillary stresses in the lung: when $\zeta$ is smaller, capillary forces are more likely to overcome the elastic energy penalty of holding lung branches shut. Both effects likely hinder the opening of the lungs during respiration, giving rise to respiratory distress in diseased patients. 
 
 Our simulations with varying $\gamma$ confirm this expectation. In particular, we find that the dynamics of lung opening strongly depend on the surface tension (Fig. \ref{Fig_gamma}a), indicating that it is another key regulator of breathing; increasing the surface tension decreases the rate at which air is drawn in and prolongs the overall duration of inhalation (Fig. \ref{Fig_gamma}c). Indeed, when $\gamma$ is increased by just a factor of two from its typical healthy value $\approx15$ mN/m \cite{geiser2003, alonso2005}, as can be the case in COVID-19 and COPD \cite{bracco2020covid, gunther2001surfactant, gralinski2015molecular, yi2020covid,kim2013chronic, hohlfeld2001role, ingenito2005role}, the full duration of lung opening takes nearly four times longer, and only reaches a tenth of the fully opened volume, as shown by the squares and circles in Fig. \ref{Fig_gamma}b, respectively. Intriguingly, the duration of inhalation varies non-monotically with $\gamma$, as shown by the squares: when $\gamma$ is small, the capillary pressure threshold $p^{th}$ is easily overcome and the lungs open quickly, while as $\gamma$ increases, capillarity increasingly resists lung opening and the duration of inhalation increases. However, as $\gamma$ increases above $\approx21$ mN/m, capillarity holds increasing numbers of branches of the lungs shut, and inhalation is truncated---causing the duration of inhalation to decrease again. This behavior also manifests in the simulated pressure-volume (Fig. \ref{Fig_gamma}d) and flow rate-volume (Fig. \ref{Fig_gamma}c, inset) curves, which are again strikingly similar to those observed in experimental measurements \cite{kozlowska2005spirometry}. Indeed, we even observe the previously-reported \cite{alencar2002dynamic, crane1973switching} non-monotonic variation of $p_{tp}$ with $V$ at low $\gamma$, as shown by the dark purple curve: under these conditions, because the capillary pressure threshold $p^{th}$ is easily overcome, rapid lung opening causes an abrupt decrease in the lung pressure---a phenomenon that has been termed an ``elastic shock'' \cite{alencar2002dynamic, crane1973switching}. Together, these results indicate that our computational approach provides a way to quantify the impact of specific changes in mucus surface tension on inhalation, isolating its relative influence in causing respiratory distress.

 \subsubsection*{Role of airway wall stiffness} The elasticity of the airway wall changes greatly in disease, often in opposing ways. For example, buildup of excess fibrous connective tissue stiffens the airway walls in CF, COPD, and asthma \cite{limjunyawong2015measurement, tiddens2010cystic, coates1930occurrence, wright2006advances}, while weakening of the tissue leads to weaker airway walls in emphysema \cite{oliveira2016entropy,suki2012mechanical, ohnishi1998matrix, de2007stress}; whether lung tissue elasticity increases, decreases, or stays unchanged is currently still being studied for COVID-19. The analysis presented in the \textit{Theory} section suggests that weakening of the airway walls in emphysema hinders lung opening during inhalation, and is likely the main contributor to respiratory distress in this case: capillary forces due to the surface tension of the mucus lining tend to hold the soft walls of closed branches together. Conversely, we expect that stiffening of the airway walls in CF, COPD, and asthma paradoxically promotes lung opening, in opposition to the respiratory distress associated with these conditions: stiffer lungs are more difficult to bend and close shut. Thus, in these cases, we expect that respiratory distress arises instead due to changes in other biomechanical factors, such as mucus viscosity and surface tension, as suggested by clinical studies \cite{puchelle2002airway, rogers2006treatment, piatti2005effects}. 

 This competition between lung elasticity and capillarity is quantified by the biomechanical parameter $\zeta\equiv \frac{R_{17}}{L_{17}}\frac{E\left(T_{17}/R_{17}\right)^{3}}{12\left(1-\nu^2\right)\left(\gamma/R_{17}\right)}$. When $\zeta$ is large, elastic stresses, as quantified by the characteristic bending stiffness $B_{17}\equiv E\left(T_{17}/R_{17}\right)^{3}\bigg/12\left(1-\nu^2\right)$, dominate and peel the lung branches open; conversely, while when $\zeta$ is small, capillarity, as quantified by the characteristic capillary pressure $\gamma/R_{17}$, dominates and tends to hold the lung branches shut. Our simulations with varying $\zeta$, exploring the full physiological range of $\zeta$ using measurements of the variation that arises in $E$ and $\gamma$ \cite{weibel1963, wang2011, lewis2001mechanics, baconnais1999, puchelle1983, geiser2003, alonso2005}, confirm this expectation. Similar to the cases of varying $\sigma_{0}$ and $\gamma$, the dynamics of lung opening strongly depend on $\zeta$ (Fig. \ref{Fig_zeta}a), indicating that it is another key regulator of breathing; decreasing $\zeta$ decreases the rate at which air is drawn in and prolongs the overall duration of inhalation (Fig. \ref{Fig_zeta}c). Intriguingly, similar to the case of $\gamma$, the duration of inhalation varies non-monotically with $\zeta$, as shown by the squares in Fig. \ref{Fig_zeta}b: when $\zeta$ is large, the lungs open quickly due to elastic stresses, while as $\zeta$ decreases, elastic stresses do not pull the lungs open as quickly and the duration of inhalation increases. However, as $\zeta$ decreases below $\approx10^{-4}$, elastic stresses cannot open many branches of the lungs, and inhalation is again truncated. This behavior also manifests in the simulated pressure-volume (Fig. \ref{Fig_zeta}d) and flow rate-volume (Fig. \ref{Fig_zeta}c, inset) curves, which are again strikingly similar to those observed in experimental measurements \cite{barreiro2004, gold2016}.

 We expect that because $E$ decreases in emphysema \cite{oliveira2016entropy,suki2012mechanical, ohnishi1998matrix, de2007stress}, and $\gamma$ possibly increases \cite{kim2013chronic, hohlfeld2001role, ingenito2005role}, $\zeta$ concurrently decreases and is thus the key biomechanical parameter that controls the onset of respiratory distress in this case. Conversely, because $E$ increases in CF, COPD, and asthma \cite{limjunyawong2015measurement, tiddens2010cystic, coates1930occurrence, wright2006advances}, $\zeta$ may not decrease in these cases---suggesting that the onset of respiratory distress is instead controlled by increases in the mucus viscosity or surface tension, as described above. Thus, our computational approach provides a way to separately quantify the impact of specific changes in airway wall stiffness $E$ on breathing, shedding light on its relative influence in causing respiratory distress.

\section*{Discussion}

The work described here represents a first step toward developing a model of the lungs that accurately describes the multi-scaled spatial and temporal features of respiration, while still managing to be computationally tractable. Our dynamic network approach explicitly resolves the relevant length and time scales of branch opening during inhalation, while also capturing how opening propagates through the interconnected and hierarchical architecture of the lungs. We demonstrate this principle by directly connecting alterations in four key biomechanical factors---the strength of thoracic muscle contraction, the viscosity and surface tension of the airway mucus lining, and the elasticity of the airway wall---to overall alterations in breathing, in qualitative agreement with experimental and clinical findings. Our model thus helps to establish how lung biomechanics impact respiration---both deepening our fundamental understanding of this ubiquitous process, and helping to elucidate how disease-induced changes in tissue-scale factors give rise to respiratory distress. However, despite the similarity between our results and published measurements, direct validation against systematic experimental measurements or more sophisticated models, for which the values of all the input parameters are known, will be a crucial next step.

Given the increasing prevalence of respiratory diseases \cite{respiratory2017}, there is a critical need for computational tools capable of quantitatively assessing the efficacy of different therapeutic interventions, such as mechanical ventilation, exogenous administration of lung surfactant, and exogenous administration of mucus thinners. The work described here addresses this critical need. Specifically, it yields a generally-applicable computational model for which measured treatment-induced changes in biomechanical parameters---e.g. $\{\sigma_{0},\gamma,\zeta,\mu,K_{ip}\}$---can be input, and the impact on breathing outcome can be assessed. Because different treatments alter lung biomechanics in different ways, this approach may yield useful insights into how treatments that influence these specific parameters will affect breathing dynamics in general---and may eventually provide a straightforward way to quickly assess the impact of different treatments for a given patient.



The model presented here focused on the case of inhalation starting from a completely closed respiratory zone as a proof of principle; however, the dynamics described in the \textit{Theory} section can be extended in future work to also describe the closure of individual lung branches due to compression of the thoracic cavity, as well as breathing dynamics in a lung with regional atelectasis, involving a mixture of both open, partially-closed, and fully-closed branches. Accomplishing this extension will require development of a form of Eq. \ref{eq2.8} that characterizes branch closure instead of opening. Further, for both inhalation and exhalation, Eq. \ref{eq2.8} can be replaced by the results of more sophisticated tube models that incorporate non-axisymmetric deformation modes, possible collapse of the mucus film, mucus-wall liquid-solid interactions arising from the competition between viscous stress, capillary stresses, and wall deformations, and heterogeneities in airway branch geometry \cite{horsfield1971models, hazel2003, heil2008, juel_heap_2007, heil1999airway, hazel2005surface, heil1997stokes, heil1996large, moriarty1999flow, romano2019liquid, grotberg2011respiratory, halpern1992fluid, grotberg1994pulmonary, grotberg2004biofluid, heil2011fluid, gaver1996steady,mauroy2015toward}. Finally, while our network representation necessarily simplifies many of the rich complexities of the lung in favor of ease of computation, it can be extended by incorporating different lung architectures, by directly inputting specific values of $\{R_{ij},L_{ij},T_{ij}\}$; heterogeneity in the biomechanical parameter values, by directly inputting specific values of $\{E_{ij},\gamma_{ij},\mu_{ij}\}$; and non-Newtonian mucus rheology, by incorporating a rate-dependent viscosity in Eq. \ref{eq2.8}. Exploring the influence of these different features on breathing will be an important extension of our work.

\begin{figure*}[t]
     \centering
         {\includegraphics[width=\textwidth]{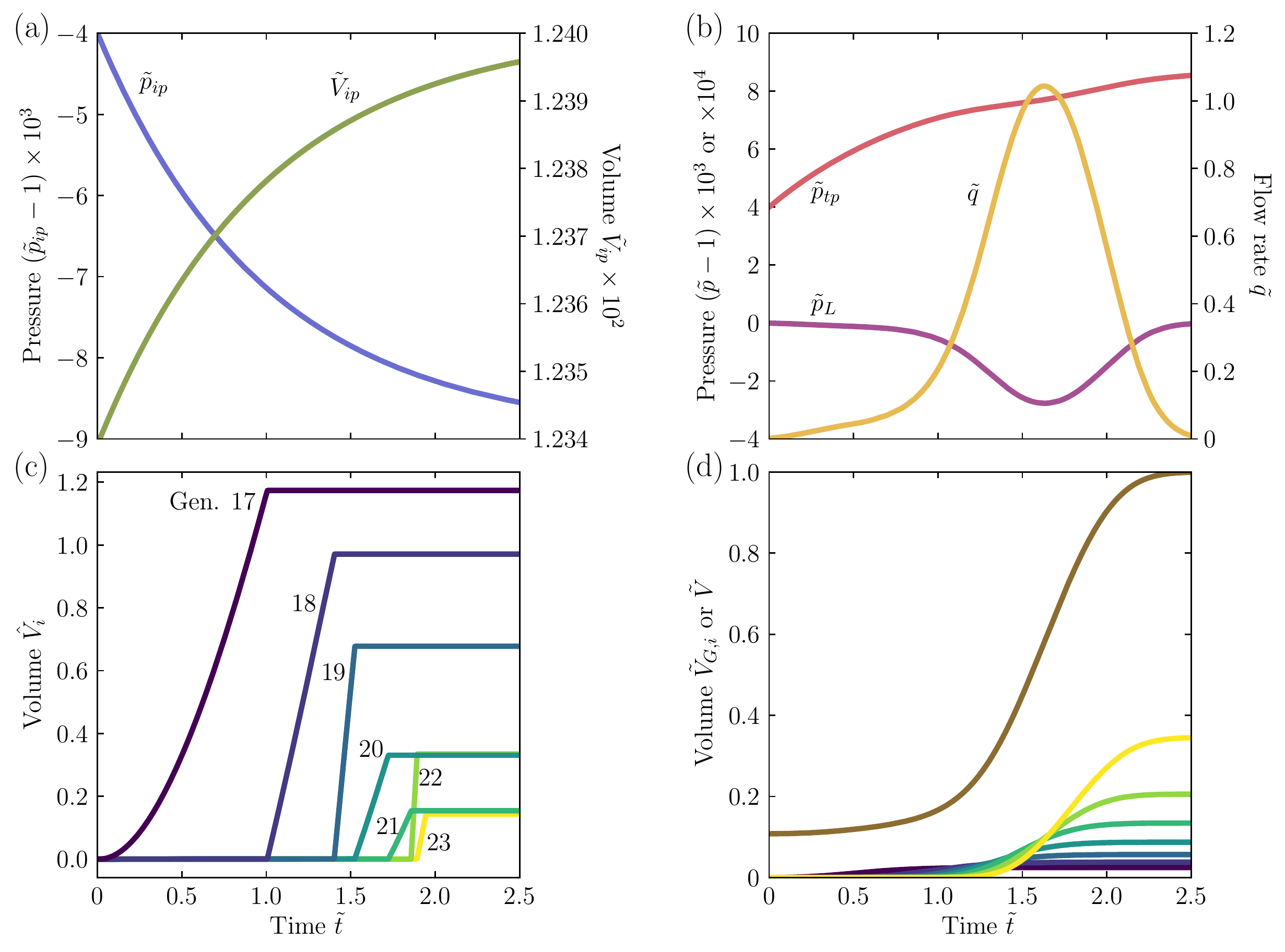}}\\
      \caption{Simulation results do not change with finer discretization. Plots show the simulated evolution of lung pressures and volumes just as in Fig. \ref{Fig2}, but with a smaller time step $\Delta\tilde{t}=10^{-4}$; results are indistinguishable from those presented in the main text with a large time step $\Delta\tilde{t}=10^{-3}$.
}\label{Fig6}
 \end{figure*}

\section*{Methods}
To simulate the dynamics of inhalation, we implement the rules given in the \textit{Theory} section in discretized form, evaluating volumes, pressures, and flow rates at successive time steps separated by $\Delta \tilde{t}$ using the iterative scheme described below. We use two different notations to differentiate between branch-scale quantities and overall lung-scale quantities. The tilde notation $(\tilde{})$ indicates that pressure, volume, flow rate, time, and flow resistance have been normalized by the atmospheric pressure $p_{0}$, maximal open lung volume $V_{0}$, characteristic flow rate $V_{0}/\tau_{B}$, breathing time $\tau_{B}$, and characteristic flow resistance $p_{0}\tau_{B}/V_{0}$, respectively. The hat notation $(\hat{})$ indicates that the variables $V_{ij}$, $t$, $R_{ij}$, $\Gamma_{ij}$, and $p_{tp}-p_{ij}^{th}$ have been normalized by the characteristic branch-scale quantities $\pi R_{17}^{2} L_{17}$, $\mu/p_{17}^{th}$, $\left(\gamma/R_{17}\right)/B_{17}$, and $p_{17}^{th}$, respectively, where the subscript 17 refers to the mean value at the first generation of the respiratory zone. For simplicity, we assume that the air is an ideal gas at a fixed temperature. 


\begin{enumerate}

\item The applied stress $\sigma_{0}$ forces the volume of the intrapleural cavity to increase:
\begin{equation}
  \tilde{V}_{ip}\left(\tilde{t} + \Delta \tilde{t}\right) =  \tilde{V}_{ip}\left(\tilde{t}\right)+ \tilde{V}_{ip,0} \left[\varepsilon\left(\tilde{t}+\Delta \tilde{t}\right) - \varepsilon\left(\tilde{t}\right)\right] \,,
  \label{eq3}
\end{equation}
\noindent where $\varepsilon(t)$ is given by Eq. \ref{eqn:epsilon}.

\item Given a fixed amount of air within the intrapleural space, the expansion of the intrapleural cavity causes the pressure in the intrapleural cavity to concomitantly decrease:
\begin{equation}
  \tilde{p}_{ip}\left(\tilde{t} + \Delta \tilde{t}\right) =   \frac{\tilde{p}_{ip}\left(\tilde{t}\right)\tilde{V}_{ip}\left(\tilde{t}\right)}{\tilde{V}_{ip}\left(\tilde{t}+\Delta \tilde{t}\right)} \,.
\end{equation}

\item This decrease in intrapleural pressure transiently increases the transpulmonary pressure, which we estimate as:
\begin{equation}
  \tilde{p}_{tp}\left(\tilde{t} + \Delta \tilde{t}\right) \approx \tilde{p}_L\left(\tilde{t}\right) - \tilde{p}_{ip}\left(\tilde{t} + \Delta \tilde{t}\right)\,.
\end{equation}
We take $\tilde{p}_L$ to be a constant throughout the respiratory zone due to the low air flow resistance of the respiratory zone. In particular, we compare the flow resistance of air through the conducting zone or through the respiratory zone, $\Omega_{C}\approx\sum_{i=0}^{16} \left( \sum_{j=1}^{2^{i}} \Omega_{ij}^{-1} \right)^{-1}$ or $\Omega_{R}\approx\sum_{i=17}^{23} \left( \sum_{j=1}^{2^{i}} \Omega_{ij}^{-1} \right)^{-1}$, respectively, where the individual branch flow resistance $\Omega_{ij}=8 \mu_{air} L_{i}/\pi R_{i}^4$ is given by the Hagen-Poiseuille equation with an air viscosity $\mu_{air}\approx18.5~\mu$Pa-s. Using measurements of $L_{i}$ and $R_{i}$ throughout the airways \cite{weibel1963, habib1994}, we estimate $\Omega_{C}\approx16.4$ Pa-s/L and $\Omega_{R}\approx0.2$ Pa-s/L. Since $\Omega_{R}\ll\Omega_{C}$, we assume that the air flow resistance of the lungs is given by that of the conducting zone, and the air pressure is constant throughout the respiratory zone.

\item For each branch $ij$ that is in contact with the open region of the lungs, if $p_{tp}$ exceeds the threshold $p_{ij}^{th}$, this pressure difference across the branch wall forces it to open, as given by Eq. \ref{eq2.8}: 
\begin{equation}
  \hat{V}_{ij}\left(\tilde{t} + \Delta \tilde{t}\right) = \hat{V}_{ij}\left(\tilde{t}\right) + \frac{ \zeta \hat{R}_{ij}^3}{\hat{\Gamma}_{ij}}
  \left[\hat{p}_{tp}(\tilde{t}+\Delta \tilde{t}) - \hat{p}_{ij}^{th} \right]\Delta \tilde{t}\,.
\end{equation}

\item As branches open, the open volume of the lungs $V=V_{C}+\sum_{i=17}^{23}\sum_{j=1}^{2^{i}}V_{ij}$ increases, causing the pressure in the respiratory zone $p_{L}$ to transiently decrease to an intermediate value $p_{L,int}$. In normal respiration, the lungs are an open system, and air in the lungs can be treated as incompressible, so that its density does not vary with the pressure changes that arise during breathing due to inertial and viscous losses in the airways. However, in our discretized representation of the lungs, for sufficiently small $\Delta \tilde{t}$, the lungs can be approximated to be a closed system at each intermediate time step: the time scale of volume changes of the lungs is much shorter than the characteristic air inflow time scale. Thus, assuming a constant $V$ and a constant amount of air within the respiratory zone during this intermediate step, we estimate the intermediate pressure as:
\begin{equation}
    \tilde{p}_{L,int}(\tilde{t}+\Delta \tilde{t}) = 
    \frac{\tilde{p}_{L}(\tilde{t}) \tilde{V}_{L}(\tilde{t})}{\tilde{V}_{L}(\tilde{t} + \Delta \tilde{t})}\,.
    \label{boyle1}
\end{equation}

\item This decrease in pressure draws air into the lungs from the atmosphere with a volumetric flow rate $q$, driven by the pressure difference $\Delta p_{int}\equiv p_{0}-p_{L,int}$. To evaluate this flow rate, we first consider the limit of $\Omega_{C}\rightarrow0$; in this case, the pressure difference is fully equilibrated at each time step, and conservation of the amount of air exchanged yields $q=\frac{\Delta p_{int} V}{p_{0}\Delta t}=\frac{\Delta p_{int}}{\Omega_{0}}$, where $\Omega_{0}\equiv p_{0}\Delta t/V$ is an intrinsic resistance that reflects the discrete time formulation of the simulation. For the case of $\Omega_{C}>0$, we then modify this expression to also incorporate $\Omega_{C}$:
\begin{equation}
  \tilde{q}(\tilde{t}+\Delta \tilde{t}) = \frac{\Delta \tilde{p}_{int}(\tilde{t}+\Delta \tilde{t})}{\tilde{\Omega}_{0}(\tilde{t}+\Delta \tilde{t})+\tilde{\Omega}_{C} }.
      \label{boyle2}
\end{equation}

\item Because $\Omega_{C}>0$, this air flow does not fully equilibrate the pressure difference $\Delta p_{int}$. Instead, the pressure in the respiratory zone at the end of the time step is given by:

\begin{equation}
  \tilde{p}_{L}(\tilde{t} + \Delta \tilde{t}) =  \tilde{p}_{L,int}(\tilde{t} + \Delta \tilde{t}) + 
  \frac{\tilde{q}(\tilde{t} + \Delta \tilde{t})\Delta \tilde{t}}{\tilde{V}_{L}(\tilde{t}+ \Delta \tilde{t})} \,.
   \label{eq9}
\end{equation}

\end{enumerate}

\noindent For each simulation presented in the main text, we iteratively solve Eqs. \ref{eq3}--\ref{eq9} over successive time steps separated by $\Delta\tilde{t}=10^{-3}$ up to $\tilde{t}=20$. We obtain identical results with even finer discretization, as shown for the case of $\Delta\tilde{t}=10^{-4}$ in Fig. \ref{Fig6}, validating the assumptions made in Steps 5--6 above. This iterative solving is done using a C++ framework that explicitly considers a given lung network structure described by the input morphological parameters $\{ R_{ij},L_{ij},T_{ij},V_{ip,0} \}$ and the biomechanical parameters $\{ \mu,\gamma,E,\nu,\sigma_{0},K_{ip},p_{ip,0} \}$. This framework is split into a layer of virtual classes that treat memory management, multithreading, and provide the basic functions to create a branched network; each simulation is then derived from these virtual base classes with data structures defining the specific parameters that are input into the model. This scheme thus enables us to determine the full evolution of the pressures $\{\tilde{p}_{ip},\tilde{p}_{tp},\tilde{p}_{L}\}$, the volumes $\{\tilde{V}_{ip},\hat{V}_{ij},\tilde{V}\}$ and the flow rate $\tilde{q}$ over time $\tilde{t}$.

\subsection*{Author Contributions}
All authors helped to design simulations; S.S.D. and J-F.L. designed and performed the theoretical analysis; F.K. performed all simulations; J-F.L., F.K.,
and S.S.D. analyzed the data; J-F.L., F.K., and S.S.D. discussed the results and implications and wrote the manuscript; S.S.D.
designed and supervised the overall project.

\subsection*{Acknowledgments}
This work was supported by startup funds from
Princeton University as well as partial support from the Keller Center REACH program for F.K. It is a pleasure to acknowledge Nathanael Ji and Anvitha Sudhakar for their work on a preliminary version of the network model at the inception of this project. 

\subsection*{Data Availability}
The code necessary to reproduce the results reported
here and to further explore the dynamic network model will be available in a Github repository at \url{https://github.com/FelixKratz/LungFramework}.

\nocite{*}




\begin{thebibliography}{0}%
\makeatletter
\providecommand \@ifxundefined [1]{%
 \@ifx{#1\undefined}
}%
\providecommand \@ifnum [1]{%
 \ifnum #1\expandafter \@firstoftwo
 \else \expandafter \@secondoftwo
 \fi
}%
\providecommand \@ifx [1]{%
 \ifx #1\expandafter \@firstoftwo
 \else \expandafter \@secondoftwo
 \fi
}%
\providecommand \natexlab [1]{#1}%
\providecommand \enquote  [1]{``#1''}%
\providecommand \bibnamefont  [1]{#1}%
\providecommand \bibfnamefont [1]{#1}%
\providecommand \citenamefont [1]{#1}%
\providecommand \href@noop [0]{\@secondoftwo}%
\providecommand \href [0]{\begingroup \@sanitize@url \@href}%
\providecommand \@href[1]{\@@startlink{#1}\@@href}%
\providecommand \@@href[1]{\endgroup#1\@@endlink}%
\providecommand \@sanitize@url [0]{\catcode `\\12\catcode `\$12\catcode
  `\&12\catcode `\#12\catcode `\^12\catcode `\_12\catcode `\%12\relax}%
\providecommand \@@startlink[1]{}%
\providecommand \@@endlink[0]{}%
\providecommand \url  [0]{\begingroup\@sanitize@url \@url }%
\providecommand \@url [1]{\endgroup\@href {#1}{\urlprefix }}%
\providecommand \urlprefix  [0]{URL }%
\providecommand \Eprint [0]{\href }%
\providecommand \doibase [0]{https://doi.org/}%
\providecommand \selectlanguage [0]{\@gobble}%
\providecommand \bibinfo  [0]{\@secondoftwo}%
\providecommand \bibfield  [0]{\@secondoftwo}%
\providecommand \translation [1]{[#1]}%
\providecommand \BibitemOpen [0]{}%
\providecommand \bibitemStop [0]{}%
\providecommand \bibitemNoStop [0]{.\EOS\space}%
\providecommand \EOS [0]{\spacefactor3000\relax}%
\providecommand \BibitemShut  [1]{\csname bibitem#1\endcsname}%
\let\auto@bib@innerbib\@empty
\end{thebibliography}%


%


\begin{thebibliography}{111}

\bibitem{xu2020pathological}
Zhe Xu, Lei Shi, Yijin Wang, Jiyuan Zhang, Lei Huang, Chao Zhang, Shuhong Liu,
  Peng Zhao, Hongxia Liu, Li~Zhu, et~al.
\newblock Pathological findings of {COVID-19} associated with acute respiratory
  distress syndrome.
\newblock {\em The Lancet respiratory medicine}, 8(4):420--422, 2020.

\bibitem{sohrabi2020world}
Catrin Sohrabi, Zaid Alsafi, Niamh O’Neill, Mehdi Khan, Ahmed Kerwan, Ahmed
  Al-Jabir, Christos Iosifidis, and Riaz Agha.
\newblock World health organization declares global emergency: A review of the
  2019 novel coronavirus ({COVID-19}).
\newblock {\em International Journal of Surgery}, 2020.

\bibitem{qin2020dysregulation}
Chuan Qin, Luoqi Zhou, Ziwei Hu, Shuoqi Zhang, Sheng Yang, Yu~Tao, Cuihong Xie,
  Ke~Ma, Ke~Shang, Wei Wang, et~al.
\newblock Dysregulation of immune response in patients with {COVID-19} in
  wuhan, china.
\newblock {\em Clinical Infectious Diseases}, 2020.

\bibitem{poyiadji2020covid}
Neo Poyiadji, Gassan Shahin, Daniel Noujaim, Michael Stone, Suresh Patel, and
  Brent Griffith.
\newblock {COVID-19}--associated acute hemorrhagic necrotizing encephalopathy:
  {CT} and {MRI} features.
\newblock {\em Radiology}, page 201187, 2020.

\bibitem{chen2020sars}
Wen-Hsiang Chen, Ulrich Strych, Peter~J Hotez, and Maria~Elena Bottazzi.
\newblock The {SARS-CoV-2} vaccine pipeline: an overview.
\newblock {\em Current tropical medicine reports}, pages 1--4, 2020.

\bibitem{xu2020clinical}
Xiao-Wei Xu, Xiao-Xin Wu, Xian-Gao Jiang, Kai-Jin Xu, Ling-Jun Ying, Chun-Lian
  Ma, Shi-Bo Li, Hua-Ying Wang, Sheng Zhang, Hai-Nv Gao, et~al.
\newblock Clinical findings in a group of patients infected with the 2019 novel
  coronavirus ({SARS-Cov-2}) outside of wuhan, china: retrospective case
  series.
\newblock {\em bmj}, 368, 2020.

\bibitem{luo2020clinical}
Weiren Luo, Hong Yu, Jizhou Gou, Xiaoxing Li, Yan Sun, Jinxiu Li, and Lei Liu.
\newblock Clinical pathology of critical patient with novel coronavirus
  pneumonia ({COVID-19}).
\newblock {\em Preprints}, 2020:2020020407, 2020.

\bibitem{polin2014surfactant}
Richard~A Polin, Waldemar~A Carlo, et~al.
\newblock Surfactant replacement therapy for preterm and term neonates with
  respiratory distress.
\newblock {\em Pediatrics}, 133(1):156--163, 2014.

\bibitem{mac2016acute}
Rob Mac~Sweeney and Daniel~F McAuley.
\newblock Acute respiratory distress syndrome.
\newblock {\em The Lancet}, 388(10058):2416--2430, 2016.

\bibitem{thompson2017acute}
B~Taylor Thompson, Rachel~C Chambers, and Kathleen~D Liu.
\newblock Acute respiratory distress syndrome.
\newblock {\em New England Journal of Medicine}, 377(6):562--572, 2017.

\bibitem{oliveira2016entropy}
Cl{\'a}udio~LN Oliveira, Asc{\^a}nio~D Ara{\'u}jo, Jason~HT Bates, Jos{\'e}~S
  Andrade~Jr, and B{\'e}la Suki.
\newblock Entropy production and the pressure--volume curve of the lung.
\newblock {\em Frontiers in Physiology}, 7:73, 2016.

\bibitem{robinson2002mucociliary}
Michael Robinson and Peter~TB Bye.
\newblock Mucociliary clearance in cystic fibrosis.
\newblock {\em Pediatric pulmonology}, 33(4):293--306, 2002.

\bibitem{newhouse1998intrapulmonary}
Patricia~A Newhouse, Fred White, John~H Marks, and Douglas~N Homnick.
\newblock The intrapulmonary percussive ventilator and flutter device compared
  to standard chest physiotherapy in patients with cystic fibrosis.
\newblock {\em Clinical pediatrics}, 37(7):427--432, 1998.

\bibitem{davis1978assisted}
Pamela~B Davis and Paul~A di~Sant'Agnese.
\newblock Assisted ventilation for patients with cystic fibrosis.
\newblock {\em Jama}, 239(18):1851--1854, 1978.

\bibitem{fauroux1999chest}
Brigitte Fauroux, Mich{\`e}le Boul{\'e}, Fr{\'e}d{\'e}ric Lofaso,
  Fran{\c{c}}oise Z{\'e}rah, Annick Cl{\'e}ment, Alain Harf, and Daniel Isabey.
\newblock Chest physiotherapy in cystic fibrosis: improved tolerance with nasal
  pressure support ventilation.
\newblock {\em Pediatrics}, 103(3):e32--e32, 1999.

\bibitem{mcelvaney1991aerosol}
Noel~G McElvaney, RC~Hubbard, P~Birrer, RG~Crystal, MS~Chernick, MM~Frank, and
  DB~Caplan.
\newblock Aerosol $\alpha$1-antitrypsin treatment for cystic fibrosis.
\newblock {\em The Lancet}, 337(8738):392--394, 1991.

\bibitem{griese1997nebulization}
Matthias Griese, P~Bufler, J~Teller, and D~Reinhardt.
\newblock Nebulization of a bovine surfactant in cystic fibrosis: a pilot
  study.
\newblock {\em European Respiratory Journal}, 10(9):1989--1994, 1997.

\bibitem{devendra2002lung}
Gehan Devendra and Roger~G Spragg.
\newblock Lung surfactant in subacute pulmonary disease.
\newblock {\em Respiratory research}, 3(1):11, 2002.

\bibitem{clark2003potential}
H~Clark and K~Reid.
\newblock The potential of recombinant surfactant protein d therapy to reduce
  inflammation in neonatal chronic lung disease, cystic fibrosis, and
  emphysema.
\newblock {\em Archives of disease in childhood}, 88(11):981--984, 2003.

\bibitem{griese1997pulmonary}
Matthias Griese, P~Birrer, and A~Demirsoy.
\newblock Pulmonary surfactant in cystic fibrosis.
\newblock {\em European Respiratory Journal}, 10(9):1983--1988, 1997.

\bibitem{homnick1995comparison}
Douglas~N Homnick, Fred White, and Carol de~Castro.
\newblock Comparison of effects of an intrapulmonary percussive ventilator to
  standard aerosol and chest physiotherapy in treatment of cystic fibrosis.
\newblock {\em Pediatric pulmonology}, 20(1):50--55, 1995.

\bibitem{holland2003non}
Anne~E Holland, Linda Denehy, G~Ntoumenopoulos, Matthew~T Naughton, and John~W
  Wilson.
\newblock Non-invasive ventilation assists chest physiotherapy in adults with
  acute exacerbations of cystic fibrosis.
\newblock {\em Thorax}, 58(10):880--884, 2003.

\bibitem{hodson1991non}
ME~Hodson, BP~Madden, MH~Steven, VT~Tsang, and MH~Yacoub.
\newblock Non-invasive mechanical ventilation for cystic fibrosis patients--a
  potential bridge to transplantation.
\newblock {\em European Respiratory Journal}, 4(5):524--527, 1991.

\bibitem{walters2011efficient}
D~Keith Walters, Greg~W Burgreen, David~M Lavallee, David~S Thompson, and
  Robert~L Hester.
\newblock Efficient, physiologically realistic lung airflow simulations.
\newblock {\em IEEE Transactions on Biomedical Engineering}, 58(10):3016--3019,
  2011.

\bibitem{soni2013large}
Bela Soni and Shahrouz Aliabadi.
\newblock Large-scale cfd simulations of airflow and particle deposition in
  lung airway.
\newblock {\em Computers \& Fluids}, 88:804--812, 2013.

\bibitem{ma2006anatomically}
Baoshun Ma and Kenneth~R Lutchen.
\newblock An anatomically based hybrid computational model of the human lung
  and its application to low frequency oscillatory mechanics.
\newblock {\em Annals of biomedical engineering}, 34(11):1691--1704, 2006.

\bibitem{wall2010towards}
Wolfgang~A Wall, Lena Wiechert, Andrew Comerford, and Sophie Rausch.
\newblock Towards a comprehensive computational model for the respiratory
  system.
\newblock {\em International Journal for Numerical Methods in Biomedical
  Engineering}, 26(7):807--827, 2010.

\bibitem{lewis2005quantification}
Tina~A Lewis, Yang-Sheng Tzeng, Erin~L McKinstry, Angela~C Tooker, Kwansoo
  Hong, Yanping Sun, Joey Mansour, Zachary Handler, and Mitchell~S Albert.
\newblock Quantification of airway diameters and {3D} airway tree rendering
  from dynamic hyperpolarized {3He} magnetic resonance imaging.
\newblock {\em Magnetic Resonance in Medicine: An Official Journal of the
  International Society for Magnetic Resonance in Medicine}, 53(2):474--478,
  2005.

\bibitem{nowak2003computational}
Natalya Nowak, Prashant~P Kakade, and Ananth~V Annapragada.
\newblock Computational fluid dynamics simulation of airflow and aerosol
  deposition in human lungs.
\newblock {\em Annals of biomedical engineering}, 31(4):374--390, 2003.

\bibitem{sznitman2007cfd}
J~Sznitman, S~Schmuki, R~Sutter, A~Tsuda, and Thomas R{\"o}sgen.
\newblock Cfd investigation of respiratory flows in a space-filling pulmonary
  acinus model.
\newblock {\em Modelling in Medicine and Biology VII, WIT Transactions on
  Biomedicine and Health}, 12:147--156, 2007.

\bibitem{walters2011computational}
D~Keith Walters and William~H Luke.
\newblock Computational fluid dynamics simulations of particle deposition in
  large-scale, multigenerational lung models.
\newblock {\em Journal of biomechanical engineering}, 133(1), 2011.

\bibitem{walters2010method}
D~Keith Walters and William~H Luke.
\newblock A method for three-dimensional navier--stokes simulations of
  large-scale regions of the human lung airway.
\newblock {\em Journal of Fluids Engineering}, 132(5), 2010.

\bibitem{kunz2009progress}
Robert~Francis Kunz, Daniel~C Haworth, DP~Porzio, and Andres Kriete.
\newblock Progress towards a medical image through cfd analysis toolkit for
  respiratory function assessment on a clinical time scale.
\newblock In {\em 2009 IEEE International Symposium on Biomedical Imaging: From
  Nano to Macro}, pages 382--385. IEEE, 2009.

\bibitem{calay2002numerical}
Rajnish~Kaur Calay, Jutarat Kurujareon, and Arne~Erik Hold{\o}.
\newblock Numerical simulation of respiratory flow patterns within human lung.
\newblock {\em Respiratory physiology \& neurobiology}, 130(2):201--221, 2002.

\bibitem{malve2013cfd}
M~Malve, S~Chandra, JL~Lopez-Villalobos, EA~Finol, A~Ginel, and M~Doblare.
\newblock Cfd analysis of the human airways under impedance-based boundary
  conditions: application to healthy, diseased and stented trachea.
\newblock {\em Computer methods in biomechanics and biomedical engineering},
  16(2):198--216, 2013.

\bibitem{wall2008fluid}
Wolfgang~A Wall and Timon Rabczuk.
\newblock Fluid--structure interaction in lower airways of {CT}-based lung
  geometries.
\newblock {\em International Journal for Numerical Methods in Fluids},
  57(5):653--675, 2008.

\bibitem{gemci2008computational}
T~Gemci, Valery Ponyavin, Y~Chen, H~Chen, and R~Collins.
\newblock Computational model of airflow in upper 17 generations of human
  respiratory tract.
\newblock {\em Journal of Biomechanics}, 41(9):2047--2054, 2008.

\bibitem{hazel2003}
Andrew~L Hazel and Matthias Heil.
\newblock Three-dimensional airway reopening: the steady propagation of a
  semi-infinite bubble into a buckled elastic tube.
\newblock {\em Journal of Fluid Mechanics}, 478:47--70, 2003.

\bibitem{heil2008}
Matthias Heil, Andrew~L Hazel, and Jaclyn~A Smith.
\newblock The mechanics of airway closure.
\newblock {\em Respiratory physiology \& neurobiology}, 163(1-3):214--221,
  2008.

\bibitem{grotberg2011respiratory}
James~B Grotberg.
\newblock Respiratory fluid mechanics.
\newblock {\em Physics of Fluids}, 23(2):021301, 2011.

\bibitem{halpern1992fluid}
D~Halpern and JB~Grotberg.
\newblock Fluid-elastic instabilities of liquid-lined flexible tubes.
\newblock {\em Journal of Fluid Mechanics}, 244:615--632, 1992.

\bibitem{grotberg1994pulmonary}
JB~Grotberg.
\newblock Pulmonary flow and transport phenomena.
\newblock {\em Annual Review of Fluid Mechanics}, 26(1):529--571, 1994.

\bibitem{grotberg2004biofluid}
James~B Grotberg and Oliver~E Jensen.
\newblock Biofluid mechanics in flexible tubes.
\newblock {\em Annual review of fluid mechanics}, 36, 2004.

\bibitem{heil2011fluid}
Matthias Heil and Andrew~L Hazel.
\newblock Fluid-structure interaction in internal physiological flows.
\newblock {\em Annual review of fluid mechanics}, 43:141--162, 2011.

\bibitem{gaver1996steady}
Donald~P Gaver, David Halpern, Oliver~E Jensen, and James~B Grotberg.
\newblock The steady motion of a semi-infinite bubble through a flexible-walled
  channel.
\newblock {\em Journal of Fluid Mechanics}, 319:25--65, 1996.

\bibitem{hazel2008influence}
Andrew~L Hazel and Matthias Heil.
\newblock The influence of gravity on the steady propagation of a semi-infinite
  bubble into a flexible channel.
\newblock {\em Physics of Fluids}, 20(9):092109, 2008.

\bibitem{juel_heap_2007}
Anne Juel and Alexandra Heap.
\newblock The reopening of a collapsed fluid-filled elastic tube.
\newblock {\em Journal of Fluid Mechanics}, 572:287–310, 2007.

\bibitem{heil1999airway}
Matthias Heil.
\newblock Airway closure: occluding liquid bridges in strongly buckled elastic
  tubes.
\newblock {\em Journal of Biomechanical Engineering}, 121(5):487--493, 1999.

\bibitem{hazel2005surface}
Andrew~L Hazel and Matthias Heil.
\newblock Surface-tension-induced buckling of liquid-lined elastic tubes: a
  model for pulmonary airway closure.
\newblock {\em Proceedings of the Royal Society A: Mathematical, Physical and
  Engineering Sciences}, 461(2058):1847--1868, 2005.

\bibitem{heil1997stokes}
Matthias Heil.
\newblock Stokes flow in collapsible tubes: computation and experiment.
\newblock {\em Journal of Fluid Mechanics}, 353(1):285--312, 1997.

\bibitem{heil1996large}
Matthias Heil and Tim~J Pedley.
\newblock Large post-buckling deformations of cylindrical shells conveying
  viscous flow.
\newblock {\em Journal of Fluids and Structures}, 10(6):565--599, 1996.

\bibitem{wongviriyawong2010dynamics}
Chanikarn Wongviriyawong, Tilo Winkler, R~Scott Harris, and Jose~G Venegas.
\newblock Dynamics of tidal volume and ventilation heterogeneity under
  pressure-controlled ventilation during bronchoconstriction: a simulation
  study.
\newblock {\em Journal of Applied Physiology}, 109(4):1211--1218, 2010.

\bibitem{politi2010multiscale}
Antonio~Z Politi, Graham~M Donovan, Merryn~H Tawhai, Michael~J Sanderson,
  Anne-Marie Lauzon, Jason~HT Bates, and James Sneyd.
\newblock A multiscale, spatially distributed model of asthmatic airway
  hyper-responsiveness.
\newblock {\em Journal of theoretical biology}, 266(4):614--624, 2010.

\bibitem{donovan2017inter}
Graham~M Donovan.
\newblock Inter-airway structural heterogeneity interacts with dynamic
  heterogeneity to determine lung function and flow patterns in both asthmatic
  and control simulated lungs.
\newblock {\em Journal of theoretical biology}, 435:98--105, 2017.

\bibitem{stewart2015patterns}
Peter~S Stewart and Oliver~E Jensen.
\newblock Patterns of recruitment and injury in a heterogeneous airway network
  model.
\newblock {\em Journal of The Royal Society Interface}, 12(111):20150523, 2015.

\bibitem{weibel1963}
Ewald~R. Weibel.
\newblock {\em Morphometry of the Human Lung}.
\newblock Academic PressInc., Publishers, 1963.

\bibitem{horsfield1971models}
Keith Horsfield, Gladys Dart, Dan~E Olson, Giles~F Filley, and Gordon Cumming.
\newblock Models of the human bronchial tree.
\newblock {\em Journal of applied physiology}, 31(2):207--217, 1971.

\bibitem{nucci2002morphometric}
Gianluca Nucci, Simonluca Tessarin, and Claudio Cobelli.
\newblock A morphometric model of lung mechanics for time-domain analysis of
  alveolar pressures during mechanical ventilation.
\newblock {\em Annals of Biomedical Engineering}, 30(4):537--545, 2002.

\bibitem{habib1994}
Robert~H Habib, Richard~B Chalker, B{\'e}la Suki, and Andrew~C Jackson.
\newblock Airway geometry and wall mechanical properties estimated from
  subglottal input impedance in humans.
\newblock {\em Journal of Applied Physiology}, 77(1):441--451, 1994.

\bibitem{wang2011}
Jau-Yi Wang, Patrick Mesquida, Prathap Pallai, Chris~J Corrigan, and Tak~H Lee.
\newblock Dynamic properties of human bronchial airway tissues.
\newblock {\em arXiv preprint arXiv:1111.5645}, 2011.

\bibitem{lewis2001mechanics}
Mark~A Lewis and Markus~R Owen.
\newblock The mechanics of lung tissue under high-frequency ventilation.
\newblock {\em SIAM Journal on Applied Mathematics}, 61(5):1731--1761, 2001.

\bibitem{laifook1978}
S.~J. Lai-Fook, R.~E. Hyatt, and J.~R. Rodarte.
\newblock Elastic constants of trapped lung parenchyma.
\newblock {\em Journal of Applied Physiology}, 44(6):853--858, 1978.
\newblock PMID: 670008.

\bibitem{baconnais1999}
Sonia Baconnais, Rabindra Tirouvanziam, Jean-Marie Zahm, Sophie de~Bentzmann,
  Bruno P{\'e}ault, G{\'e}rard Balossier, and Edith Puchelle.
\newblock Ion composition and rheology of airway liquid from cystic fibrosis
  fetal tracheal xenografts.
\newblock {\em American journal of respiratory cell and molecular biology},
  20(4):605--611, 1999.

\bibitem{puchelle1983}
E~Puchelle, JM~Zahm, and C~Duvivier.
\newblock Spinability of bronchial mucus. relationship with viscoelasticity and
  mucous transport properties.
\newblock {\em Biorheology}, 20(2):239--249, 1983.

\bibitem{matsui2006}
Hirotoshi Matsui, Victoria~E Wagner, David~B Hill, Ute~E Schwab, Troy~D Rogers,
  Brian Button, Russell~M Taylor, Richard Superfine, Michael Rubinstein,
  Barbara~H Iglewski, et~al.
\newblock A physical linkage between cystic fibrosis airway surface dehydration
  and pseudomonas aeruginosa biofilms.
\newblock {\em Proceedings of the National Academy of Sciences},
  103(48):18131--18136, 2006.

\bibitem{hughes1999}
Philip~D Hughes, Michael~I Polkey, M~Lou~Harris, Andrew~JS Coats, John Moxham,
  and Malcolm Green.
\newblock Diaphragm strength in chronic heart failure.
\newblock {\em American journal of respiratory and critical care medicine},
  160(2):529--534, 1999.

\bibitem{bland1986}
J~Martin Bland and Douglas~G Altman.
\newblock Statistical methods for assessing agreement between two methods of
  clinical measurement.
\newblock {\em The lancet}, 327(8476):307--310, 1986.

\bibitem{ker1998}
J~Ker et~al.
\newblock Respiratory muscle endurance in heart failure--the effect of clinical
  severity.
\newblock {\em Cardiovascular Journal of Africa}, 9(1):20--23, 1998.

\bibitem{polkey1996}
Michael~I Polkey, Dimitris Kyroussis, Carl-Hugo Hamnegard, Gary~H Mills,
  Malcolm Green, and John Moxham.
\newblock Diaphragm strength in chronic obstructive pulmonary disease.
\newblock {\em American journal of respiratory and critical care medicine},
  154(5):1310--1317, 1996.

\bibitem{d2019physiology}
Horacio~P D'Agostino and Mary~Ann Edens.
\newblock Physiology, pleural fluid.
\newblock In {\em StatPearls [Internet]}. StatPearls Publishing, 2019.

\bibitem{christie1934}
Ronald~V Christie and CA~McIntosh.
\newblock The measurement of the intrapleural pressure in man and its
  significance.
\newblock {\em Journal of Clinical Investigation}, 13(2):279, 1934.

\bibitem{geiser2003}
Marianne Geiser, Samuel Schurch, and Peter Gehr.
\newblock Influence of surface chemistry and topography of particles on their
  immersion into the lung's surface-lining layer.
\newblock {\em Journal of applied physiology}, 94(5):1793--1801, 2003.

\bibitem{alonso2005}
Coralie Alonso, Alan Waring, and Joseph~A Zasadzinski.
\newblock Keeping lung surfactant where it belongs: protein regulation of
  two-dimensional viscosity.
\newblock {\em Biophysical journal}, 89(1):266--273, 2005.

\bibitem{ellwein2018theoretical}
Laura Ellwein~Fix, Joseph Khoury, Russell~R Moores~Jr, Lauren Linkous, Matthew
  Brandes, and Henry~J Rozycki.
\newblock Theoretical open-loop model of respiratory mechanics in the extremely
  preterm infant.
\newblock {\em PloS one}, 13(6):e0198425, 2018.

\bibitem{zhang2013dynamic}
Xiangming Zhang and Rong~Z Gan.
\newblock Dynamic properties of human tympanic membrane based on
  frequency-temperature superposition.
\newblock {\em Annals of biomedical engineering}, 41(1):205--214, 2013.

\bibitem{blom2003}
JA~Blom.
\newblock {\em Monitoring of respiration and circulation}.
\newblock CRC Press, 2003.

\bibitem{suki1994avalanches}
B{\'e}la Suki, Albert-L{\'a}szl{\'o} Barab{\'a}si, Zolt{\'a}n Hantos, Ferenc
  Pet{\'a}k, and H~Eugene Stanley.
\newblock Avalanches and power-law behaviour in lung inflation.
\newblock {\em Nature}, 368(6472):615--618, 1994.

\bibitem{suki2000size}
B{\'e}la Suki, Adriano~M Alencar, J{\'o}zsef Tolnai, Tibor Asztalos, Ferenc
  Pet{\'a}k, Mamatha~K Sujeer, Keena Patel, Jignish Patel, H~Eugene Stanley,
  and Zolt{\'a}n Hantos.
\newblock Size distribution of recruited alveolar volumes in airway reopening.
\newblock {\em Journal of Applied Physiology}, 89(5):2030--2040, 2000.

\bibitem{bates2002time}
Jason~HT Bates and Charles~G Irvin.
\newblock Time dependence of recruitment and derecruitment in the lung: a
  theoretical model.
\newblock {\em Journal of Applied Physiology}, 93(2):705--713, 2002.

\bibitem{wang2020updated}
Weier Wang, Jianming Tang, and Fangqiang Wei.
\newblock Updated understanding of the outbreak of 2019 novel coronavirus
  ({2019-nCoV}) in wuhan, china.
\newblock {\em Journal of medical virology}, 92(4):441--447, 2020.

\bibitem{kim2013chronic}
Victor Kim and Gerard~J Criner.
\newblock Chronic bronchitis and chronic obstructive pulmonary disease.
\newblock {\em American journal of respiratory and critical care medicine},
  187(3):228--237, 2013.

\bibitem{tobin2010narrative}
Martin~J Tobin, Franco Laghi, and Amal Jubran.
\newblock Narrative review: ventilator-induced respiratory muscle weakness.
\newblock {\em Annals of internal medicine}, 153(4):240--245, 2010.

\bibitem{kozlowska2005spirometry}
Wanda~J Kozlowska and Paul Aurora.
\newblock Spirometry in the pre-school age group.
\newblock {\em Paediatric respiratory reviews}, 6(4):267--272, 2005.

\bibitem{lai2009micro}
Samuel~K Lai, Ying-Ying Wang, Denis Wirtz, and Justin Hanes.
\newblock Micro-and macrorheology of mucus.
\newblock {\em Advanced drug delivery reviews}, 61(2):86--100, 2009.

\bibitem{rubin2007mucus}
Bruce~K Rubin.
\newblock Mucus structure and properties in cystic fibrosis.
\newblock {\em Paediatric respiratory reviews}, 8(1):4--7, 2007.

\bibitem{sturgess1970viscosity}
Jennifer~M Sturgess, AJ~Palfrey, and Lynne Reid.
\newblock The viscosity of bronchial secretion.
\newblock {\em Clinical science}, 38(1):145--156, 1970.

\bibitem{ye2020chest}
Zheng Ye, Yun Zhang, Yi~Wang, Zixiang Huang, and Bin Song.
\newblock Chest {CT} manifestations of new coronavirus disease 2019
  ({COVID-19}): a pictorial review.
\newblock {\em European radiology}, pages 1--9, 2020.

\bibitem{bracco2020covid}
MD~Bracco~Lorenzo.
\newblock {COVID-19}, Type II alveolar cells and surfactant.
\newblock {\em J Med--Clin Res \& Rev}, 4(4):1--3, 2020.

\bibitem{gunther2001surfactant}
Andreas G{\"u}nther, Clemens Ruppert, Reinhold Schmidt, Philipp Markart,
  Friedrich Grimminger, Dieter Walmrath, and Werner Seeger.
\newblock Surfactant alteration and replacement in acute respiratory distress
  syndrome.
\newblock {\em Respiratory research}, 2(6):353, 2001.

\bibitem{gralinski2015molecular}
Lisa~E Gralinski and Ralph~S Baric.
\newblock Molecular pathology of emerging coronavirus infections.
\newblock {\em The Journal of pathology}, 235(2):185--195, 2015.

\bibitem{yi2020covid}
Ye~Yi, Philip~NP Lagniton, Sen Ye, Enqin Li, and Ren-He Xu.
\newblock {COVID-19}: what has been learned and to be learned about the novel
  coronavirus disease.
\newblock {\em International journal of biological sciences}, 16(10):1753,
  2020.

\bibitem{hohlfeld2001role}
Jens~M Hohlfeld.
\newblock The role of surfactant in asthma.
\newblock {\em Respiratory research}, 3(1):1--8, 2001.

\bibitem{ingenito2005role}
Edward~P Ingenito, Larry~W Tsai, Arnab Majumdar, and Bela Suki.
\newblock On the role of surface tension in the pathophysiology of emphysema.
\newblock {\em American journal of respiratory and critical care medicine},
  171(4):300--304, 2005.

\bibitem{alencar2002dynamic}
Adriano~M Alencar, Stephen~P Arold, Sergey~V Buldyrev, Arnab Majumdar,
  Dimitrije Stamenovi{\'c}, H~Eugene Stanley, and B{\'e}la Suki.
\newblock Dynamic instabilities in the inflating lung.
\newblock {\em Nature}, 417(6891):809--811, 2002.

\bibitem{crane1973switching}
HD~Crane.
\newblock Switching properties in bubbles, balloons, capillaries and alveoli.
\newblock {\em Journal of Biomechanics}, 6(4):411--422, 1973.

\bibitem{limjunyawong2015measurement}
Nathachit Limjunyawong, Jonathan Fallica, Maureen~R Horton, and Wayne Mitzner.
\newblock Measurement of the pressure-volume curve in mouse lungs.
\newblock {\em JoVE (Journal of Visualized Experiments)}, (95):e52376, 2015.

\bibitem{tiddens2010cystic}
Harm~AWM Tiddens, Scott~H Donaldson, Margaret Rosenfeld, and Peter~D Par{\'e}.
\newblock Cystic fibrosis lung disease starts in the small airways: can we
  treat it more effectively?
\newblock {\em Pediatric pulmonology}, 45(2):107--117, 2010.

\bibitem{coates1930occurrence}
George~M Coates and Matthew~S Ersner.
\newblock Occurrence of eosinophils in the mucous membrane of the maxillary
  sinus in asthmatic patients.
\newblock {\em Archives of Otolaryngology}, 11(2):158--168, 1930.

\bibitem{wright2006advances}
JL~Wright and A~Churg.
\newblock Advances in the pathology of COPD.
\newblock {\em Histopathology}, 49(1):1--9, 2006.

\bibitem{suki2012mechanical}
B{\'e}la Suki, Rajiv Jesudason, Susumu Sato, Harikrishnan Parameswaran,
  Ascanio~D Araujo, Arnab Majumdar, Philip~G Allen, and Erzs{\'e}bet
  Bartol{\'a}k-Suki.
\newblock Mechanical failure, stress redistribution, elastase activity and
  binding site availability on elastin during the progression of emphysema.
\newblock {\em Pulmonary pharmacology \& therapeutics}, 25(4):268--275, 2012.

\bibitem{ohnishi1998matrix}
Keisuke Ohnishi, Michiaki Takagi, Yoshimochi Kurokawa, Susumu Satomi, and
  Yrjo~T Konttinen.
\newblock Matrix metalloproteinase-mediated extracellular matrix protein
  degradation in human pulmonary emphysema.
\newblock {\em Laboratory investigation; a journal of technical methods and
  pathology}, 78(9):1077--1087, 1998.

\bibitem{de2007stress}
Jessica de~Ryk, Jacqueline Thiesse, Eman Namati, and Geoffrey McLennan.
\newblock Stress distribution in a three dimensional, geometric alveolar sac
  under normal and emphysematous conditions.
\newblock {\em International journal of chronic obstructive pulmonary disease},
  2(1):81, 2007.

\bibitem{puchelle2002airway}
Edith Puchelle, Odile Bajolet, and Michel Ab{\'e}ly.
\newblock Airway mucus in cystic fibrosis.
\newblock {\em Paediatric respiratory reviews}, 3(2):115--119, 2002.

\bibitem{rogers2006treatment}
Duncan~F Rogers and Peter~J Barnes.
\newblock Treatment of airway mucus hypersecretion.
\newblock {\em Annals of medicine}, 38(2):116--125, 2006.

\bibitem{piatti2005effects}
Gioia Piatti, Umberto Ambrosetti, Pierachille Santus, and Luigi Allegra.
\newblock Effects of salmeterol on cilia and mucus in COPD and pneumonia
  patients.
\newblock {\em Pharmacological research}, 51(2):165--168, 2005.

\bibitem{barreiro2004}
Timothy Barreiro and Irene Perillo.
\newblock An approach to interpreting spirometry.
\newblock {\em American family physician}, 69(5):1107--1114, 2004.

\bibitem{gold2016}
Warren~M Gold and Laura~L Koth.
\newblock Pulmonary function testing.
\newblock {\em Murray and Nadel's Textbook of Respiratory Medicine}, page 407,
  2016.

\bibitem{respiratory2017}
The global impact of respiratory disease – second edition.
\newblock Forum of International Respiratory Societies, Sheffield, European
  Respiratory Society, 2017.

\bibitem{moriarty1999flow}
John~A Moriarty and James~B Grotberg.
\newblock Flow-induced instabilities of a mucus--serous bilayer.
\newblock {\em Journal of Fluid Mechanics}, 397:1--22, 1999.

\bibitem{romano2019liquid}
Francesco Roman{\`o}, Hideki Fujioka, Metin Muradoglu, and JB~Grotberg.
\newblock Liquid plug formation in an airway closure model.
\newblock {\em Physical Review Fluids}, 4(9):093103, 2019.

\bibitem{mauroy2015toward}
Benjamin Mauroy, Patrice Flaud, Dominique Pelca, Christian Fausser, Jacques
  Merckx, and Barrett~R Mitchell.
\newblock Toward the modeling of mucus draining from human lung: role of
  airways deformation on air-mucus interaction.
\newblock {\em Frontiers in physiology}, 6:214, 2015.


\end{thebibliography}
\end{document}